\documentclass[]{emulateapj}
\usepackage{apjfonts}
\usepackage{mathptmx}

\slugcomment{Accepted to the \aj, Dec 2004}

\shorttitle{\sc Dust and Ionized Gas in Nine Early-Type Galaxies}
\shortauthors{\sc Martel et al.}

\begin{document}

\title{Dust and Ionized Gas in Nine Nearby Early-Type Galaxies Imaged
with the HST Advanced Camera for Surveys\altaffilmark{1}}

\author{A.R. Martel\altaffilmark{2},
H.C. Ford\altaffilmark{2},
L.D. Bradley\altaffilmark{2},
H.D. Tran\altaffilmark{3}, 
F. Menanteau\altaffilmark{2},
Z.I. Tsvetanov\altaffilmark{2}, 
G.D. Illingworth\altaffilmark{4},
G.F. Hartig\altaffilmark{5},
M. Clampin\altaffilmark{6}}

\email{martel@pha.jhu.edu, ford@pha.jhu.edu, ldb@pha.jhu.edu,
htran@keck.hawaii.edu, menanteau@pha.jhu.edu, zlatan@pha.jhu.edu,
gdi@ucolick.org, hartig@stsci.edu, clampin@stsci.edu}


\altaffiltext{1}{Based on observations made with the NASA/ESA {\em
Hubble Space Telescope} which is operated by AURA, Inc., under NASA
contract NAS 5-26555.}

\altaffiltext{2}{Department of Physics and Astronomy, Johns Hopkins
University, 3400 North Charles Street, Baltimore, MD 21218.}

\altaffiltext{3}{W.M. Keck Observatory, 65-1120 Mamalahoa Hwy.,
Kamuela, HI 96743.}

\altaffiltext{4}{UCO/Lick Observatory, University of California, Santa
Cruz, CA 95064.}

\altaffiltext{5}{Space Telescope Science Institute, 3700 San Martin
Drive, Baltimore, MD 21218.}

\altaffiltext{6}{NASA Goddard Space Flight Center, Code 681,
Greenbelt, MD 20771.}


\begin{abstract}
   We present $V$ and $I$ continuum images and H$\alpha$+[N~II] maps
of nine early-type galaxies observed with the Wide Field Channel of
the Advanced Camera for Surveys on the Hubble Space Telescope.  Dust
and ionized gas are detected in all galaxies.  The optical nebulae are
primarily concentrated on the nuclei and extend out to radii of a few
hundred parsecs, in compact clumps, filaments, or disks.  Two
galaxies, NGC~6166 and NGC~6338, also possess diffuse, ionized
filaments on kiloparsec scales.  The ionized gas is entirely contained
within the nuclear disks of ESO~208-G021, NGC~3078, and NGC~7720.  In
the radio-loud galaxy NGC~6166, emission-line filaments are detected
along the radio lobes, possibly as a result of shock ionization.  A
wide range of ionized gas masses,
$M_g\approx7\times10^2-3\times10^6$~$M_\odot$, are calculated from
the observed fluxes.  Even in this small sample, the orientation of
the ionized material correlates well with the major or minor axis of
the galaxies, consistent with an external origin for the dust and gas.
\end{abstract}


\section{Introduction}

   Nearby, bright early-type galaxies have been the subject of
numerous imaging studies with the {\em Hubble Space Telescope} (HST),
encompassing several facets of their formation and evolution.  The
high spatial resolution of HST has provided new insights in the
properties of their evolved stellar population, such as metallicity,
mean age, radial distribution, and color-magnitude diagrams. The
surface frequencies, core sizes, colors, luminosity functions, and
radial dependence of their globular cluster populations have also been
examined in detail (e.g. Kundu \& Whitmore 2001).  Deep imaging of
their cores has revealed rich dust morphologies in a majority of them,
such as disks and filaments (van Dokkum \& Franx 1995; de Koff et
al. 2000; Tran et al. 2001), previously unresolved with ground-based
telescopes.  Moreover, early-type galaxies have been the nearly
exclusive targets for spectroscopic measurements of supermassive black
holes (BHs) with HST.

   Not surprisingly, the great majority of these programs have been
conducted in broad-band continuum filters. Comparatively little effort
has been devoted to narrow-band imaging in prominent emission lines,
although such data is crucial for the understanding of the warm gas
phase ($T\sim10^4$~K) of the interstellar medium (ISM) and its
interaction with other phases, in particular the cold dust
($T\sim30$~K) and the hot X-ray gas ($T\sim10^7$~K).

   Deep narrow-band imaging at the resolution of HST would help
elucidate the energetics and origin of the warm gas in early-type
galaxies.  Several excitation mechanisms have been postulated, such as
stellar radiation (photoionization by star-forming regions or old
post-Asymptotic Giant Branch (AGB) stars), photoionization by an
active nucleus, internal shocks, and electron conduction, but none yet
stands out as the unequivocal favorite. In fact, a combination of
mechanisms may be at play in individual galaxies.  Moreover, although
an external origin for the gas and dust from the accretion of a
gas-rich companion galaxy is usually favored, an internal origin from
mass ejection from evolved stars or cooling from the hot phase, can
not be entirely dismissed (e.g. Mathews \& Brighenti 2003).

   Here, we present new continuum and H$\alpha$+[N~II] emission-line
images of a sample of nine early-type galaxies observed with the
Advanced Camera for Surveys (ACS; Ford et al. 2002) as part of the ACS
Guaranteed-Time-Observer (GTO) program.  These galaxies have been
well-studied from the radio to the X-ray spectral regimes, as
evidenced by the abundance of references to them in the published
literature. The galaxies were observed with the goal of mapping the
distribution of their line-emitting regions for possible follow-up
spectroscopy with HST to measure their BH masses.  The spectroscopic
results of the best candidates will be presented in a future paper.

\section{Sample Selection}

\begin{deluxetable*}{l c c c c c c}
\tabletypesize{\footnotesize}
\tablewidth{0pt}
\tablenum{1}
\tablecolumns{7}
\tablecaption{The Sample of Galaxies~: General Properties}
\tablehead{
\colhead{Galaxy} & 
\colhead{$V_{Vir}$ (km sec$^{-1}$)} &  
\colhead{$B_T$} & 
\colhead{$D$ (Mpc)} &
\colhead{$M_B$} &
\colhead{$\sigma$ (km sec$^{-1}$)} &
\colhead{Type}}
\startdata  
ESO 208-G021 &   788  &  11.43  &   11    &  -18.8  &  168  & S0 \\
NGC 404      &    82  &  10.93  &   3     &  -16.5  &   54  & S0 \\
NGC 2768     &  1633  &  10.73  &   23    &  -21.1  &  205  & S0 \\
NGC 2832     &  7003  &  12.62  &  100    &  -22.4  &  310  & E  \\
NGC 3078     &  2353  &  11.75  &   33    &  -20.9  &  237  & E  \\
NGC 3226     &  1410  &  12.17  &   20    &  -19.3  &  208  & E  \\
NGC 6166 (3C 338) &  9538  &  12.66  &  140    &  -23.0  &  321  & E  \\
NGC 6338     &  8541  &  13.30  &  120    &  -22.2  &  347  & S0 \\
NGC 7720 (3C 465) &  9215  &  13.03  &  130    &  -22.6  &  396  & E  \\
\enddata
\tablecomments{The radial velocity of the galaxy corrected
for infall towards Virgo, $V_{Vir}$, and the total apparent magnitude,
$B_T$, are tabulated from The Lyon-Meudon Extragalactic Database
(HyperLEDA - http://leda.univ-lyon1.fr/; Paturel et al. 1997). The
luminosity distance $D$ and absolute magnitude $M_B$ are derived from
these quantities using $(h, \Omega_m, \Omega_\Lambda) = (0.71, 0.27,
0.73)$ (Bennett et al. 2003) except for NGC~404, for which we quote
the more likely distance of 3~Mpc (see text).  The stellar velocity
dispersions $\sigma$ are compiled from McElroy (1995) except for
NGC~6338, which comes from HyperLEDA. The morphological type,
lenticular (S0) or elliptical (E), is taken from the NASA
Extragalactic Database (NED) which is primarily based on the Third
Reference Catalogue of Bright Galaxies (RC3; de Vaucouleurs et
al. 1995).}
\end{deluxetable*}




   The galaxies were chosen with the primary aim of testing the tight
relationship between the mass $M_{BH}$ of their central black hole and
the velocity dispersion $\sigma_\ast$ of their bulge (Ferrarese \&
Merritt 2000; Gebhardt et al. 2000) at the lowest and highest possible
velocity dispersions. To be a candidate in the ACS imaging program, we
imposed two requirements.  First, ionized gas must be present in the
nuclear regions of each galaxy. For confirmation, a literature search
of published H$\alpha$+[N~II] spectra or line ratios was made.
Second, if a black hole is present with the mass predicted by the
galaxy's velocity dispersion, the black hole's sphere of influence
$R_{BH}=G\,M_{BH}\,\sigma_\ast^{-2}$, which is the region where the
black hole's gravitational potential dominates the potential of the
stars, must be resolved with the long slit of the Space Telescope
Imaging Spectrograph (STIS) on HST.

   The selected galaxies and their global properties are listed in
Table~1.  They possess a large range of stellar velocity dispersions,
$\sigma_\ast\approx50-400$~km~sec$^{-1}$, and hence, black hole masses
of $5\times10^{5}$ to $2\times10^9$~M$_\sun$ are expected.  We note
that the sample is not statistically complete and that the targets
were not selected according to any luminosity criterion.  The
selection by $\sigma_\ast$ alone resulted in a sample with a disparate
range of properties.  The blue absolute magnitudes are in the range
$-19 \lesssim M_B \lesssim -23$.  Four of the galaxies are lenticular
(S0) while the others are elliptical.  Three (NGC~404, NGC~2768, and
NGC~3226) are LINERs and two (NGC~6166 and NGC~7720) are powerful
radio galaxies hosting 3C sources.  Nuclear dust disks are found in
ESO~208-G021, NGC~3078, and NGC~7720.

\section{Observations and Reduction}

   All the galaxies were observed in Cycles~11 and 12 as part of the
ACS/GTO programs GTO-9293 and 9986 (PI~: Ford) with the Wide Field
Channel (WFC).  The imaging device of the WFC consists of two butted
$2048\times4096$ CCDs manufactured by Scientific Imaging Technologies
(SITe). All the frames were acquired in normal operation and read out
with all four amplifiers (overall read-out noise $\approx4.9$~e$^-$)
at a gain of 1~e$^-$/DN.

   The journal of observations is tabulated in Table~2.  Each galaxy
was observed for one orbit in a narrow-band filter, either F658N
(FWHM$\approx85$~\AA) or FR656N (FWHM$\approx2$\% or 130~\AA),
spanning the redshifted H$\alpha$+[N~II] emission line complex, and in
a broad-band filter (F555W or F814W), adjacent in wavelength, for
continuum subtraction.  All the broad-band exposures were limited to a
total duration of 700~sec to avoid overhead penalties in the dumping
of the on-board buffer memory and the remainder of the orbit was
dedicated to the narrow-band exposure, typically 1400-1700~sec.  The
exposures were split into at least two sub-exposures, dithered
slightly from each other for ESO~208-G021, for a more effective
rejection of cosmic ray hits and/or hot pixels.  The images were
processed with the ACS Science Analysis Pipeline (APSIS; Blakeslee et
al. 2003) at Johns Hopkins University. The spatial scale of the final,
reduced images is $0\farcs05$ pixel$^{-1}$.

   A dust map was generated for each galaxy of the sample.  The
stellar contribution in the broad-band image was first modeled with
elliptical isophotes using the {\em ellipse} task in IRAF/STSDAS.  A
rough model was then subtracted from the original image and unwanted
features such as bad pixels, dust lanes, and globular clusters were
masked and excluded in a second pass.  The ratio of the continuum
image over its isophotal model yields the dust map~: absorption
features such as dust are dark and emission features such as globular
clusters are light.  No isophotal solution could be achieved inside a
radius of $\approx1\arcsec$ for NGC~404 because of saturation in the
central pixels.  For NGC~6166, the fit was restricted to radii
$\lesssim7\arcsec$ to achieve a valid solution over the isophotes of
the masked surrounding galaxies.

\begin{deluxetable}{l l c c c c}
\tabletypesize{\footnotesize}
\tablewidth{0pt}
\tablenum{2}
\tablecolumns{4}
\tablecaption{Journal of Observations}
\tablehead{
\colhead{Galaxy} & 
\colhead{UT Date} &
\colhead{Filter} &
\colhead{Exposure (sec)}} 
\startdata  
ESO 208-G021 & 2003 Jul 24  &  F814W  & 700  \\
             &              &  F658N  & 1550 \\
NGC 404      & 2003 Jan 8   &  F814W  & 700  \\
             &              &  F658N  & 1500 \\
NGC 2768     & 2003 Jan 14  &  F814W  & 700  \\
             &              &  F658N  & 1700 \\  
NGC 2832     & 2003 Feb 17  &  F555W  & 700  \\
             &              &  FR656N & 1400 \\
NGC 3078     & 2003 Apr 12  &  F814W  & 700  \\
             &              &  F658N  & 1700 \\
NGC 3226     & 2003 Mar 8   &  F814W  & 700  \\
             &              &  F658N  & 1400 \\
NGC 6166     & 2003 May 5   &  F814W  & 700  \\
             &              &  FR656N & 1550 \\
NGC 6338     & 2003 Jan 11  &  F814W  & 700  \\
             &              &  FR656N & 1700 \\
NGC 7720     & 2003 Nov 4   &  F814W  & 700  \\
             &              &  FR656N & 1490 \\
\enddata
\end{deluxetable}


   The isophotal analysis returns several quantities of interest~: the
surface brightness profile $\mu$, the position angle (PA) of the major
axis, the ellipticity $e=1-b/a$, and B4, the coefficient of the
$\cos(4\theta)$ component of the fit normalized by the semi-major axis
and local intensity gradient.  For $\mu$, the photometric zero-points
(ZEROPT) were calculated with the {\em calcphot} task in IRAF/SYNPHOT,
giving 32.82 for F555W and 32.61 for F814W in the Vega magnitude
system ($m_{Vega}=-2.5\log({\rm counts}/{\rm sec})+{\rm
ZEROPT}_{Vega}$).  The transformation between the ACS and Johnson
filter systems is available in Sirianni et al. (2004).  B4 measures
the deviation of the isophotes from a perfect ellipse - a negative
value indicates a box-shaped isophote while a positive value, a
disk-shaped isophote (Jedrzejewski 1987).

\setcounter{figure}{0}
\begin{figure*}
\includegraphics[height=\textwidth,angle=270]{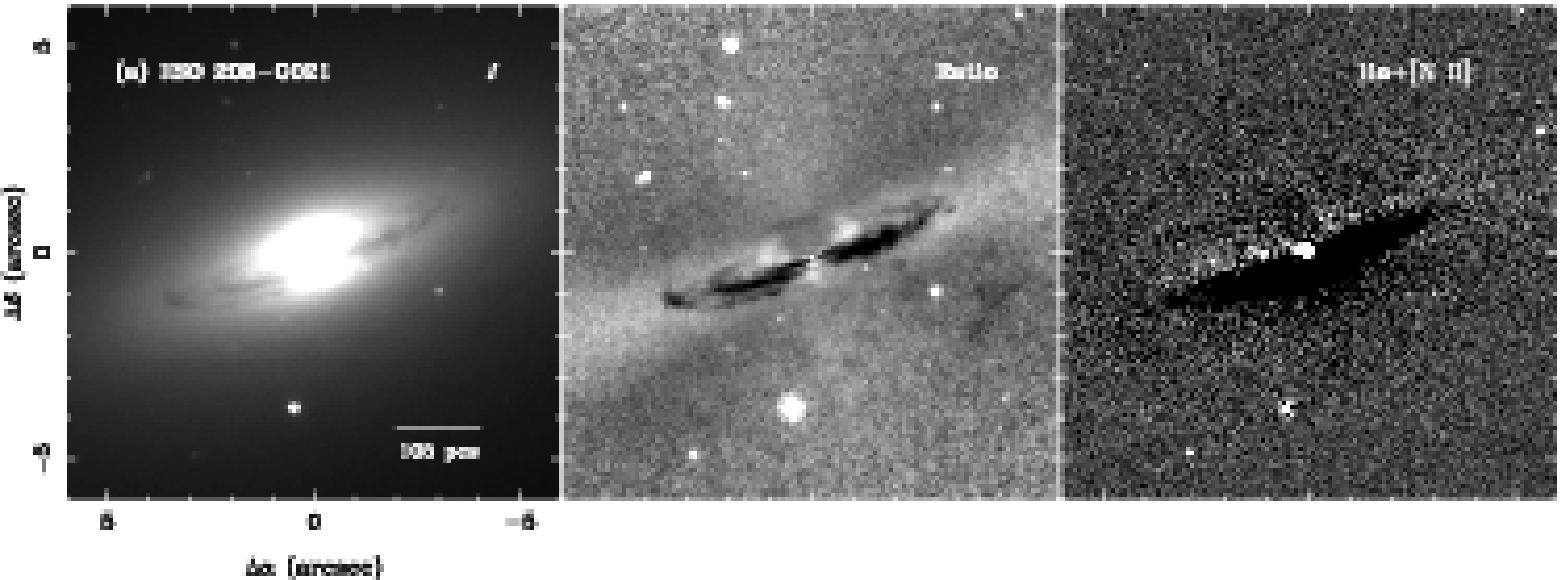}
\includegraphics[height=\textwidth,angle=270]{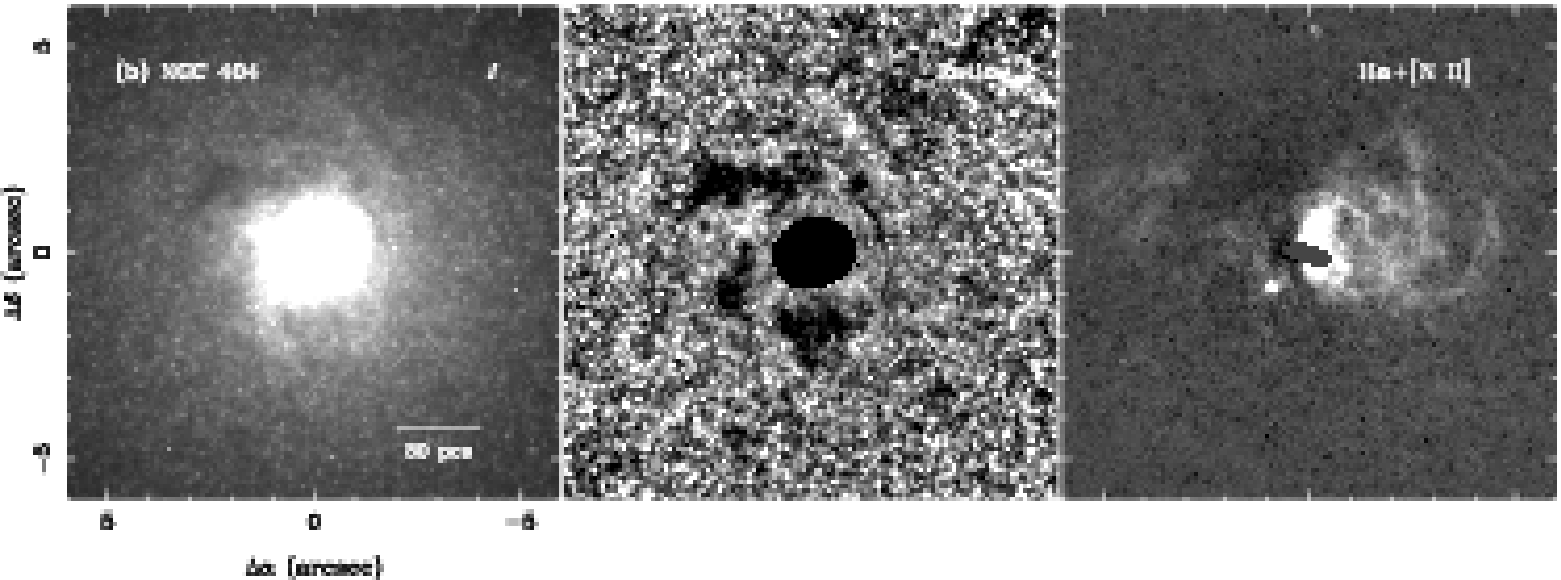}
\includegraphics[height=\textwidth,angle=270]{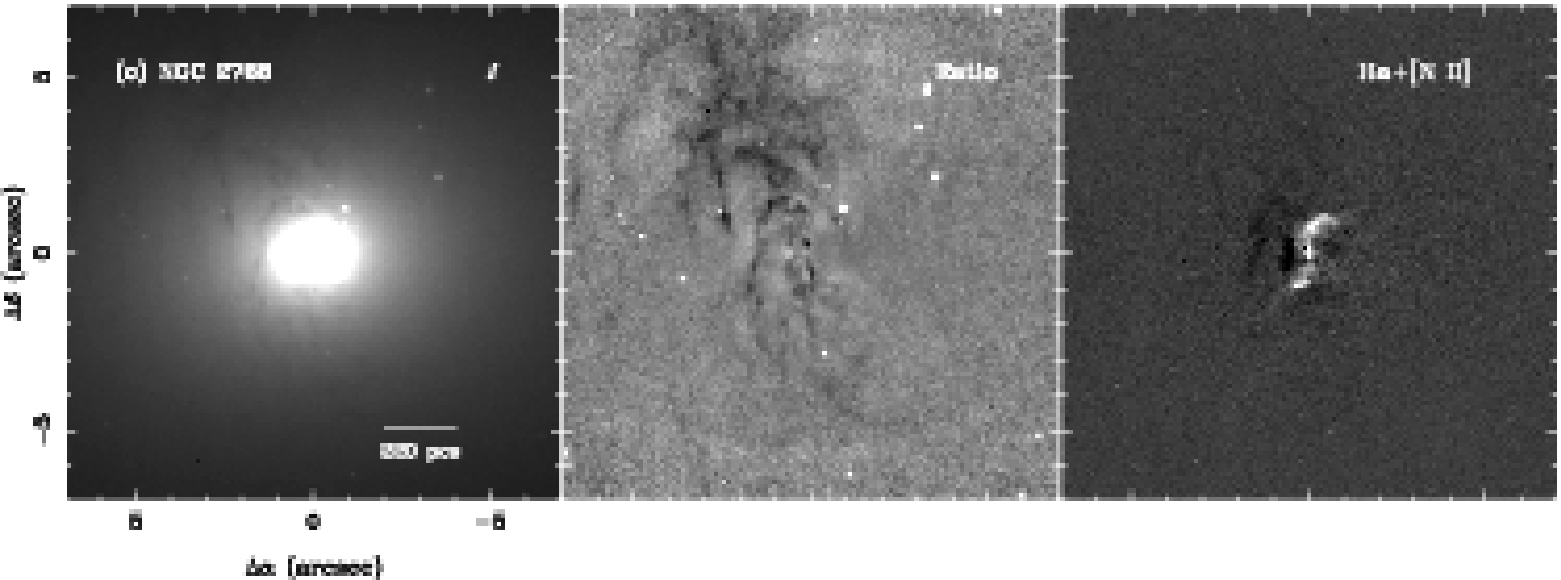}
\figcaption[martel.fig1.ps]{The broad-band image ($V$ or $I$), the
dust map, i.e. the ratio of the broad-band image over its isophotal
model, displayed between levels 0.8 and 1.2, and the H$\alpha$+[N~II]
emission-line map are displayed on a linear scale for each galaxy of
the sample.  The absorption features (dust) are dark and the emission
features (stars, globular clusters, and ionized gas) are light.  The
orientation is North-up and East-left.  In panels (d), (f), (g), and
(h), enlarged regions of the dust and ionization maps are also
shown. Some of the enlarged regions are outlined with a white dotted
line in the continuum image.  The saturated core of NGC~404 is masked
inside a radius of $\sim0\farcs9$ in the dust map, where no isophotal
fits could be successfully achieved, and along the direction of charge
``bleeding'' in the H$\alpha$+[N~II] map.  Maps of the dust and
ionized gas of NGC~2831, the neighbor of NGC~2832, are also included
in panel (d). Similarly, the core and spiral arm of NGC~3227, the
companion of NGC~3226, are enlarged in H$\alpha$+[N~II] in panel~(f).
The center of NGC~6338 is enlarged in the bottom-right panels of
panel~(h).  The direction of the radio axis of NGC~6166 (3C~338) and
NGC~7720 (3C~465) is indicated by the short line in the bottom-left
corner of their $I$-band image.}
\end{figure*}

   The extraction of the pure emission-line map went as follows.  The
intensity levels between the broad- and narrow-band images were
matched using the counts in small regions in the host galaxy.  The
scaled continuum image was then subtracted from the narrow-band image
and this offered a first approximation to the emission-line
distribution. After a close examination, this map was refined by
recalculating the scaling ratio in line- and dust-free regions.
Residual dust signatures are visible in some images, usually in the
core. This could be the result of color differences in the underlying
stellar population and to the small differential extinction between
the narrow- and broad- filters.

   The line fluxes were measured by summing the counts in regions with
positive line emission usually by delimiting the regions visually in
{\em SAOImage DS9}.  The conversion factor between the observed flux
expressed in counts~sec$^{-1}$ and the ``physical'' flux in standard
units of ergs~sec$^{-1}$~cm$^{-2}$ was calculated with the {\em
calcphot} task in IRAF/SYNPHOT by modeling the flux with a dummy
5~\AA-wide Gaussian profile at the appropriate redshift, similar to
the prescription of Biretta, Baggett, \& Noll (1996) for narrow-band
photometry with the Wide Field Planetary Camera~2 (WFPC2) on HST.  The
fluxes were then corrected for interstellar extinction using $E(B-V)$
from Schlegel, Finkbeiner, \& Davis (1998) and the Cardelli, Clayton,
\& Mathis (1989) extinction law for $R_V=3.1$, followed by a
correction for the galaxy redshift.

   The errors in the line flux measurements are largely due to the
choice of the scaling ratio and to the difficulty in defining the
edges of the regions over which to sum the counts, in particular for
the extended, low-surface brightness filaments.  To estimate the
errors, the fluxes were carefully measured on two separate occasions
and the difference was simply taken as the measurement error. We find
typical errors in a range of $5-30$\% from high- to low-surface
brightness regions. Hence, we adopt an uncertainty of 20\% on all
line fluxes.

\section{Results}
\label{results}

\subsection{Dust and Ionized Gas Morphologies}

   The continuum image, the dust map, and the continuum-subtracted
emission-line map of each galaxy of our sample are shown in Fig.~1.
The surface brightness profile $\mu$, the position angle PA, the
ellipticity $e$, and the harmonic amplitude $B4$ are plotted in
Fig.~2.  In the following, we discuss the dust and ionized gas
morphologies in more detail.

{\ \ \ } \\
\noindent
{\em ESO 208-G021 --- Fig.~1(a)~:} This galaxy possesses a nearly
edge-on nuclear disk roughly 375~pc ($7\arcsec$) in diameter and
oriented along the major axis of the galaxy ($\approx110\arcdeg$).
The ionized gas is observed along the ``northern'' edge of the disk,
suggesting that it is primarily confined to its interior. The gas
appears diffuse with a few faint compact clumps. The most prominent is
located at the geometrical center of the disk which coincides with the
galaxy center.  Scorza et al. (1998) performed a photometric
disk/bulge decomposition of ESO~208-G021 - our isophotal results are
consistent with theirs.  The B4 parameter indicates that the galaxy is
very ``disky'' throughout most of its extent.  The field around the
galaxy is pockmarked with numerous nebulosities.  A galaxy located
$2\farcm1$ to the south and first identified by Lauberts (1982) is
cleanly resolved into a nearly face-on Sc galaxy in our $I$-band
image.

{\ \ \ } \\
\noindent
{\em NGC 404 --- Fig.~1(b)~:} This extensively studied LINER galaxy,
believed to be located just outside the Local Group, possesses the
smallest stellar velocity dispersion in our sample (54~km~sec$^{-1}$).
Its distance remains a matter of debate - typical estimates have
ranged from $\approx1$~Mpc (Baars \& Wendker 1976) to $\approx10$~Mpc
(Wiklind \& Henkel 1990), but recent measurements from surface
brightness fluctuations (Jensen et al. 2003) and color-magnitude
diagrams (Tikhonov, Galazutdinova, \& Aparicio 2003) place it at a
distance of $3.0-3.5$~Mpc. We adopt the lower value of 3~Mpc.

   Previous HST images of NGC~404 in visible broad and narrow bands
have been presented in Pogge et al. (2000), Tikhonov, Galazutdinova,
\& Aparicio (2003) and Martini et al. (2003).  The greater depth of
our ACS images reveals more detail in the galaxy's circumnuclear
regions.  In our $I$-band image, individual stars are resolved. In
fact, from their WFPC2 data, Tikhonov, Galazutdinova, \& Aparicio
(2003) derived the color-magnitude diagram of NGC~404 and found that
the stellar population is typical of an early-type galaxy.  The B4
parameter shows no strong preference for boxiness or diskiness.

   As was found in the WFPC2 work, the dust and gas distributions are
complex. Three distinct, irregular clumps of dust are found in the
eastern half of the galaxy out to a radius of $\approx50$~pc
($3\farcs5$) at PAs of $50\arcdeg$, $120\arcdeg$, and $180\arcdeg$.
The overall orientation of the ionized gas is along the east-west
axis.  It consists of a bright clump of diameter $\approx35$~pc
($2\farcs4$) elongated along a north-south direction and roughly
centered on the nucleus.  Remarkably, two filaments are attached at
each end of the clump.  In projection, the filaments intertwine and
cross at a radius of $\approx30$~pc ($2\farcs1$) from the nucleus.
The filament connected to the northern end of the clump forms an
\mbox{\textsf{S}-shape} and extends the furthest to the west, out to
$\approx70$~pc ($4\farcs8$). The other filament simply curves to the
north and appears to bifurcate into two strands. Other shorter and
fainter filaments are also observed.  Pogge et al. (2000) and
Eracleous et al. (2002) interpret the gas morphology as a superbubble
blown out by a circumnuclear starburst.  In the east, the gas extends
out to $\approx130$~pc ($9\arcsec$) and is very diffuse and patchy
except for a bright, resolved knot located at the end of a filament,
at $\approx17$~pc ($1\farcs2$) SE of the nucleus.  Some of this gas is
likely heavily obscured by the dust.

\setcounter{figure}{0}
\begin{figure*}
\includegraphics[height=\textwidth,angle=270]{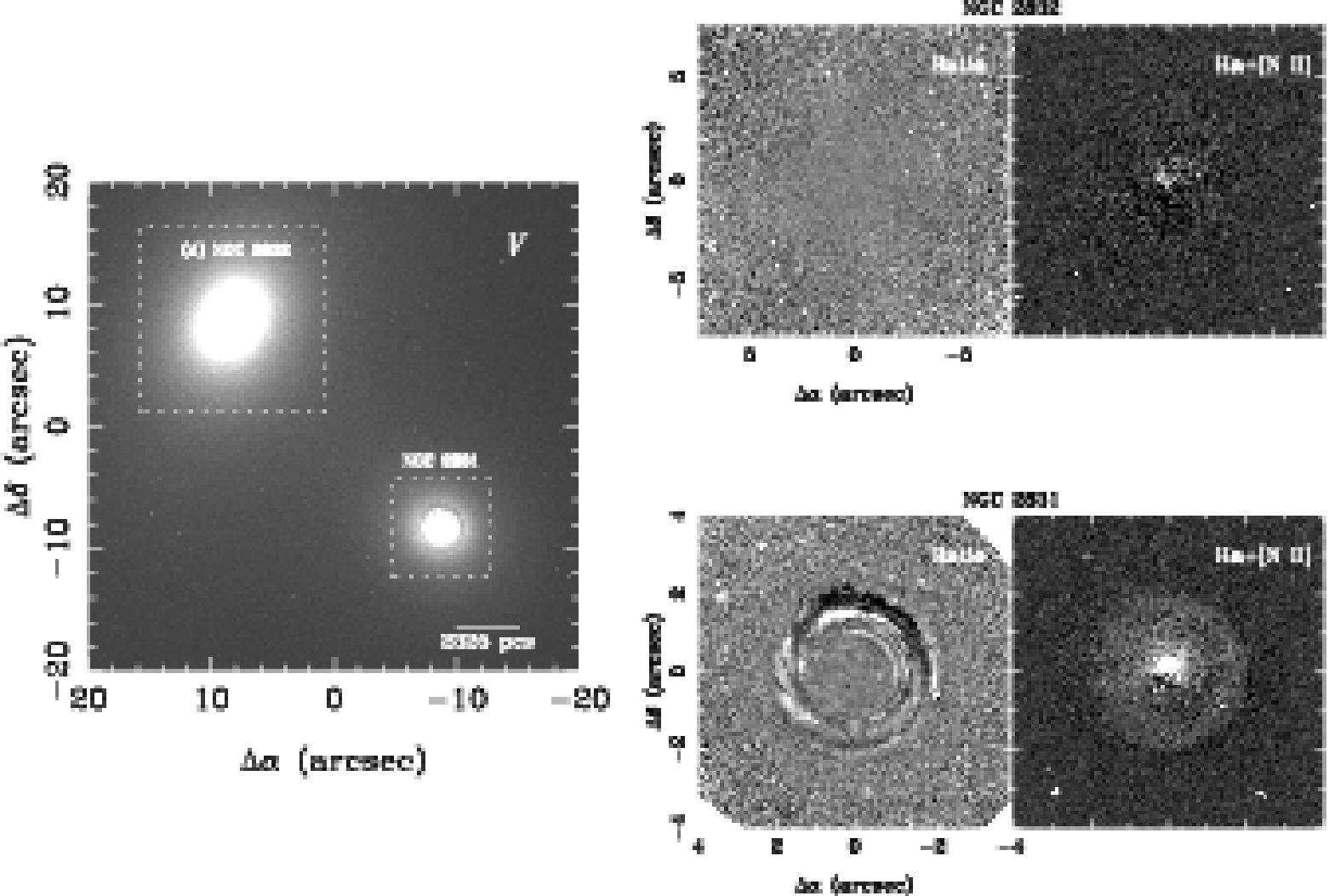}
\includegraphics[height=\textwidth,angle=270]{./martel.fig1e.ps}
\caption{{\em Continued}}
\end{figure*}

\setcounter{figure}{0}
\begin{figure*}
\includegraphics[height=\textwidth,angle=270]{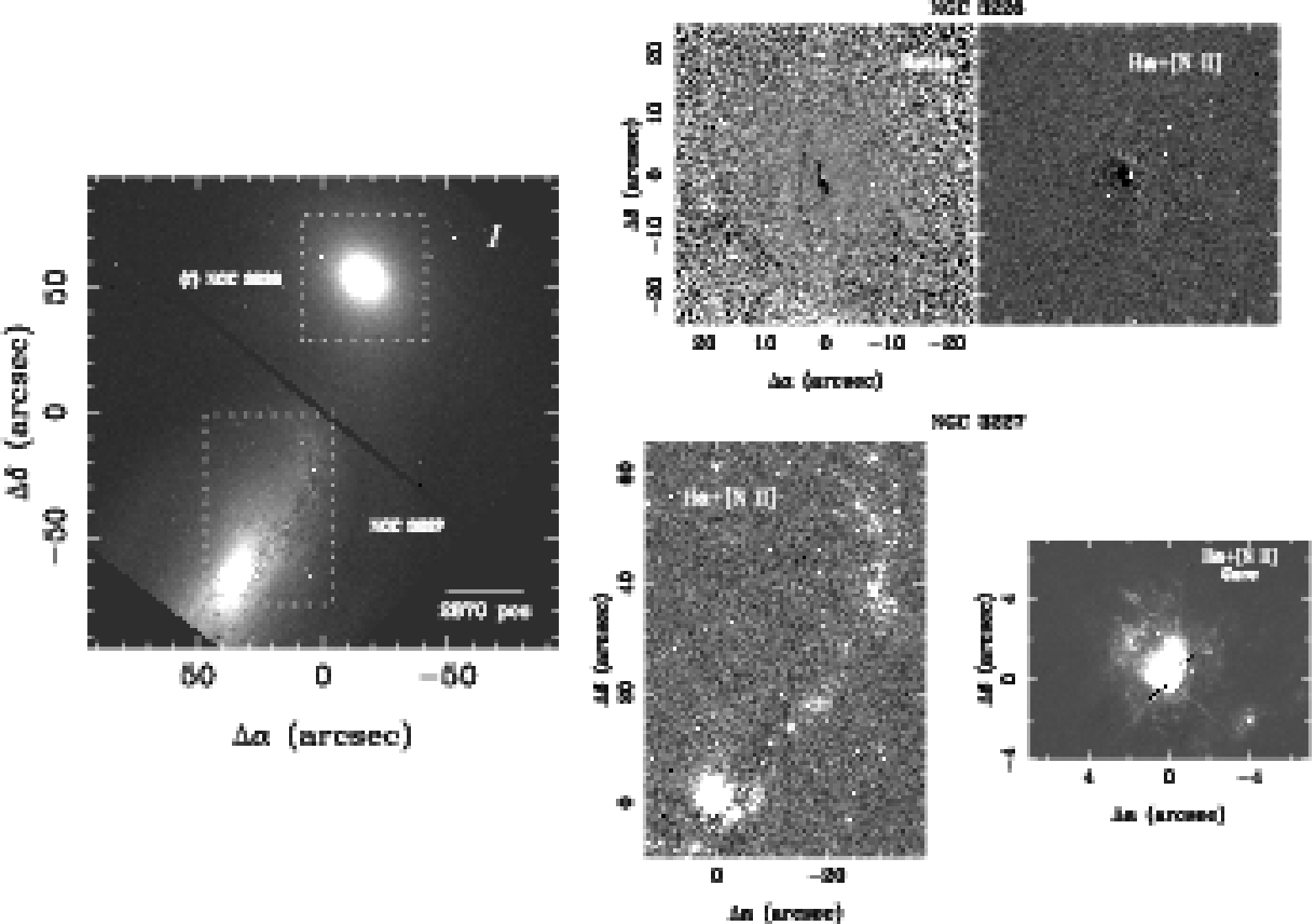}
\caption{{\em Continued}}
\end{figure*}

{\ \ \ } \\
\noindent
{\em NGC 2768 --- Fig.~1(c)~:} Another popular LINER, NGC~2768
displays a rich network of dusty filaments and knots extending to a
radius of $\approx2.4$~kpc ($22\arcsec$), mostly dominant in the northern
part of the galaxy. The main axis of the dust distribution is nearly
aligned with the minor axis of the galaxy.  The ionized gas is
concentrated in a beautiful spiral feature whose long axis is oriented
north-south. Its total length is $\approx220$~pc ($2\arcsec$) and it
appears imbedded in a diffuse halo. Each arm of the spiral consists of
a small inner clump (at radii $\lesssim45$~pc ($0\farcs4$)) completed by
a brighter curved knot.  The nucleus appears unresolved. The overall
morphologies and orientations of the gas and dust features are
consistent with the ground images presented by Kim (1989).

{\ \ \ } \\
\noindent
{\em NGC 2832 --- Fig.~1(d)~:} NGC~2832 is the cD galaxy of the
Abell~779 cluster.  Its B4 parameter indicates that the isophotes are
generally boxy.  There is little evidence of dust in the galaxy except
perhaps for a hint of a narrow, twisting lane meandering through the
core along a north-south axis. In an F814W WFPC2 image, Seppo et
al. (2003) find no signature of dust absorption in the central
$4\arcsec\times4\arcsec$.  The ionized gas consists of a clump
centered on the nucleus with a possible second and fainter clump
immediately NE of it.

   Its close companion galaxy, NGC~2831, is located $\approx11$~kpc
($24\arcsec$) to the south-west and possesses a face-on dust disk.
The redshifted H$\alpha$+[N~II] line complex of NGC~2831 also falls in
the wavelength range of the FR656N filter - its ionization map shows
that the gas is entirely confined to the interior of the disk.

{\ \ \ } \\
\noindent
{\em NGC 3078 --- Fig.~1(e)~:} An HST/WFPC2 F702W image of NGC~3078 is
shown in Tran et al. (2001) and an isophotal analysis based on this
image is presented in Rest et al. (2001).  Our ACS images clearly show
the nearly edge-on nuclear disk of diameter $\approx180$~pc
($1\farcs1$) oriented along the major axis of the galaxy.  The galaxy
isophotes are strongly boxy, in accord with Rest et al. (2001).  The
WFC field is pockmarked with a large number of unresolved globular
clusters and faint, extended galaxies.  The extraction of a reliable
emission-line map is difficult because of the presence of the disk,
but it appears that the warm gas is primarily concentrated on the
nucleus, as reported by Trinchieri \& di Serego Alighieri (1991) with
ground-based narrow-band imaging.  A small, lenticular galaxy located
$59\arcsec$ due east of NGC~3078 and identified with the infrared
source 2MASX~J09582899-2655366, is bifurcated by a thin dust lane.

\setcounter{figure}{0}
\begin{figure*}
\includegraphics[height=\textwidth,angle=270]{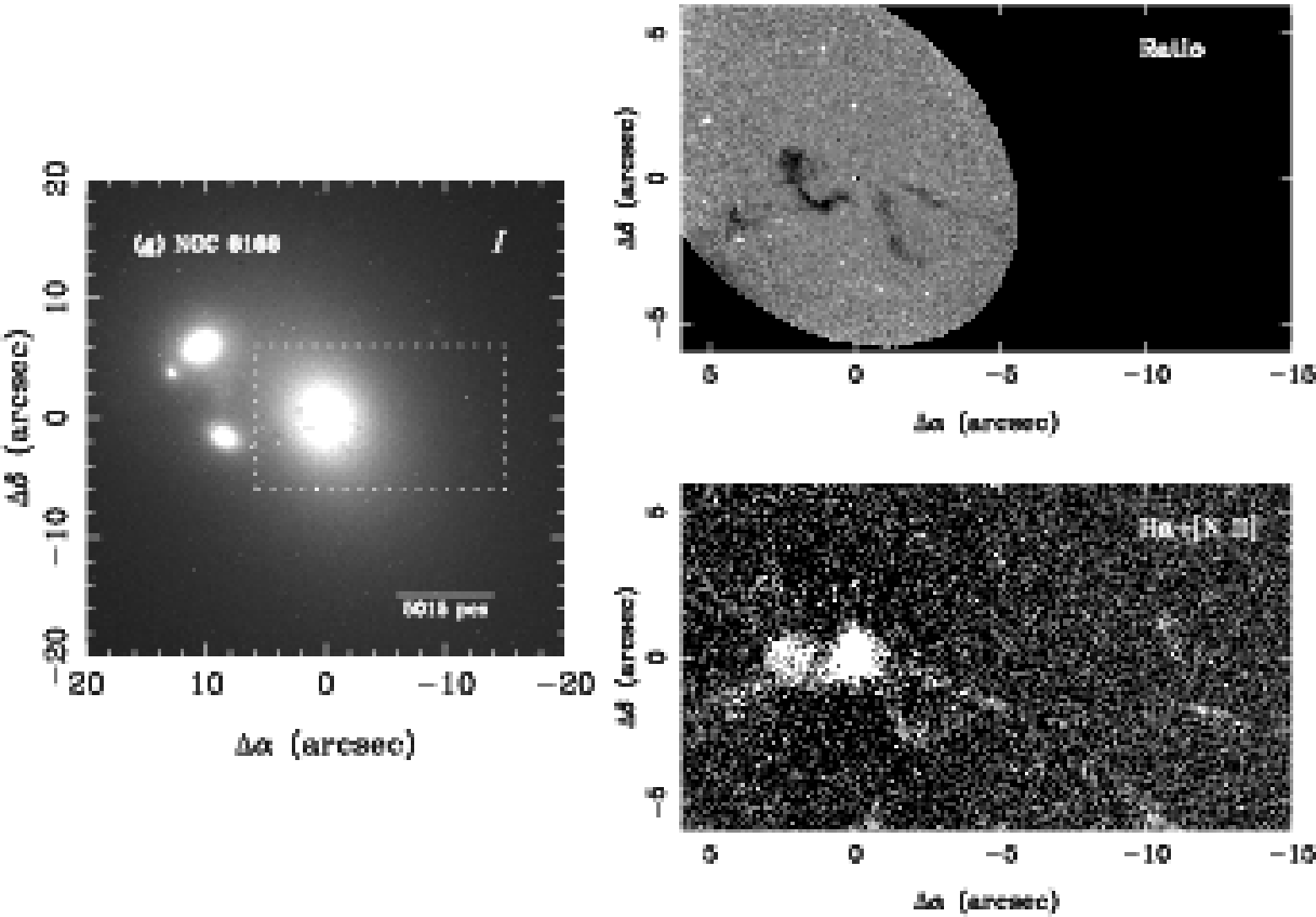}
\caption{{\em Continued}}
\end{figure*}

{\ \ \ } \\
\noindent
{\em NGC 3226 --- Fig.~1(f)~:} The dwarf elliptical LINER NGC~3226 and
its spiral neighbor NGC~3227, a popular Seyfert galaxy, form a
well-known interacting pair of galaxies.  The dust morphology of
NGC~3226 is dominated by a semi-circular lane wrapping around the
eastern side of the nucleus. Its outer radius is $\approx160$~pc
($1\farcs7$) and it lies at PAs of $20\arcdeg-200\arcdeg$.  The
northern tip of the lane is connected to a chaotic group of dusty
strands. Like the central lane, these are oriented roughly along a
north-south axis, close to the major axis of the galaxy. More knots
and filaments are detected in the south-west quadrant of the galaxy at
a radius of 1.5~kpc ($15\farcs5$) from the nucleus.  Our detailed
ionization map shows that the H$\alpha$+[N~II] gas consists primarily
of a clump of diameter 85~pc ($0\farcs9$) centered on the nucleus. Our
isophotal analysis of NGC~3226 is consistent with that of Rest et
al. (2001).

   The core as well as the northern spiral arm of NGC~3227 are also
located on our broad- and narrow-band ACS frames.  The unresolved
Seyfert nucleus is saturated in our $I$-band frame.  The spiral arm
exhibits a complicated and chaotic network of dust filaments, in
accord with the WFPC2 structure map of Pogge \& Martini (2002).  The
redshifted H$\alpha$+[N~II] lines of NGC~3227 fall within the
wavelength range of the F658N filter used to image NGC~3226. The
derived ionization map shows a large number of luminous and compact
clumps imbedded in diffuse regions in the spiral arm, likely the sites
of active star formation.  The geometry of the circumnuclear gas of
NGC~3227 has been described by Mundell et al. (1995) using
ground-based [O~III] imaging and later by Schmitt \& Kinney (1996)
using the HST Wide Field/Planetary Camera~1 (WF/PC-1).  Our ACS
continuum-subtracted maps confirm their conclusions~: a cone of
ionized gas clearly extends in the NE quadrant at an orientation of
PA$\approx30\arcdeg$. Moreover, the high spatial resolution of HST/ACS
reveals several knotty and twisting gas filaments inside the cone
radiating outward from the core out to a distance of $\approx390$~pc
($5\arcsec$).

{\ \ \ } \\
\noindent
{\em NGC 6166 --- Fig.~1(g)~:} NGC~6166, host of the powerful radio
source 3C~338, is the dominant cD galaxy in Abell~2199.  A wealth of
information on this galaxy is available in the literature over all
wavelength domains. Here, we compare our data with past ground and HST
imaging at visible wavelengths.  Continuum images acquired with HST's
WF/PC-1 and WFPC2 cameras in the F555W/$V$ and F702W/$R$ filters are
presented in Lauer et al. (1995), Martel et al. (1999), and de Koff et
al. (2000). Our new ACS images are significantly deeper and reveal
previously undetected dust and ionized filaments.

   The dust morphology consists of several filaments spread throughout
the southern half of the galaxy out to a distance of $\approx3.7$~kpc
($5\farcs8$) from the nucleus. Unfortunately, because of the nearby
galaxies that contaminate the isophotal models, the dust map can not
be extended to larger radii.  A large clump, $\approx630$~pc ($1\arcsec$)
in size, is located $\approx1.4$~kpc ($2\farcs2$) immediately east of the
nucleus, at the northern tip of a curved filament.

   Most of the emission-line flux of NGC~6166 comes from its nuclear
regions. Indeed, using ground-based narrow-band imaging, Morganti,
Ulrich, \& Tadhunter (1992) did not detect extended line emission in
this object. On the other hand, our deep ACS images reveal several
low-surface brightness filaments out to radii of $\approx8$~kpc
($13\arcsec$) in the SW quadrant. Strong line emission is also associated
with the eastern dust clump.  The strong spatial correlation on all
scales between the dust and gas features is quite remarkable in this
galaxy.

\begin{deluxetable*}{l c c c c c l}
\tabletypesize{\footnotesize}
\tablewidth{0pt}
\tablenum{3}
\tablecolumns{7}
\tablecaption{Emission-line Fluxes and Luminosities}
\tablehead{
\colhead{Galaxy} & 
\colhead{$F_{H\alpha+[N~II]}$} &
\colhead{$L_{H\alpha+[N~II]}$} &
\colhead{$F_{H\alpha}$} &
\colhead{$L_{H\alpha}$} &
\colhead{[N~II]$\lambda6583$/H$\alpha$} &
\colhead{Reference} \\
\colhead{(1)}  & 
\colhead{(2)}  & 
\colhead{(3)}  & 
\colhead{(4)}  & 
\colhead{(5)}  & 
\colhead{(6)}  & 
\colhead{(7)}}
\startdata  
ESO 208-G021 &  24.8     & 3.65     &  4.91   & 0.72    &  3.02   & Phillips et al. (1986) \\
NGC 404      &  $>149$   & $>1.60$  &  93.5   & 1.01    &  0.44   & Ho, Filippenko, \& Sargent (1997) \\
NGC 2768     &  38.4     & 24.5     &  13.0   & 8.28    &  1.46   & Ho, Filippenko, \& Sargent (1997) \\
NGC 2832     &  3.12     & 37.7     &  0.64   & 7.69    &  2.91   & Ho, Filippenko, \& Sargent (1997) \\
NGC 3078     &  \nodata  & \nodata  & \nodata & \nodata &  2.75   & Trinchieri \& di Serego Alighieri (1991)  \\
NGC 3226     &  47.9     & 22.7     &  16.3   & 7.71    &  1.45   & Ho, Filippenko, \& Sargent (1997) \\
NGC 6166     &  31.3     & 710      &  8.50   & 193     &  2.00   & Cowie et al. (1983) \\
NGC 6338     &  1.22     & 2.2      & \nodata & \nodata & \nodata & \nodata \\
NGC 7720     &  29.2     & 618      &  4.23   &  89.4   &  4.41   & de Robertis \& Yee (1990) \\
\enddata 
\tablecomments{
(1)~: Galaxy name; (2)-(3)~: $F_{H\alpha+[N~II]}$ and 
$L_{H\alpha+[N~II]}$ -- our measurements of the H$\alpha$+[N~II] line
fluxes and luminosities from the ACS/WFC narrow-band images in
$10^{-15}$~ergs~sec$^{-1}$~cm$^{-2}$ and $10^{38}$~ergs~sec$^{-1}$,
corrected for Galactic extinction and redshift. The saturated core of
NGC~404 is excluded. Uncertainties on the line fluxes are
$\approx20$\%.; (4)-(5)~: $F_{H\alpha}$ and $L_{H\alpha}$
--- the H$\alpha$ line fluxes and luminosities after removal of the
[N~II]$\lambda\lambda6548,6583$ contribution using
[N~II]$\lambda6583$/[N~II]$\lambda6548=2.94$, in the same units as
cols~(2)-3).; (6)-(7)~: the [N~II]$\lambda6583$/H$\alpha$ ratio and
its reference. For NGC~7720, the ratio with respect to the narrow
component of H$\alpha$ is listed.}
\end{deluxetable*}



{\ \ \ } \\
\noindent
{\em NGC 6338 --- Fig.~1(h)~:} NGC~6338 is the cD galaxy of a poor
cluster.  Its dust distribution consists of several knots lying along
PA$\sim160\arcdeg$.  The separation between the northernmost and
southernmost knots is $\approx3$~kpc ($5\farcs3$).  The ionized gas
consists of two luminous, compact clumps in the central arcsecond of
the galaxy and of several broken and diffuse filaments extending out
to $\approx7.5$~kpc ($13\farcs5$) in the south-east and north-west
quadrants. The southern clump coincides with the nucleus of the
galaxy. The central clumps are roughly oriented along the major axis
of the galaxy while the large-scale filaments lie along the minor
axis. Similar to NGC~6166, the dust knots trace the gas morphology
very well.

\setcounter{figure}{0}
\begin{figure*}
\includegraphics[height=\textwidth,angle=270]{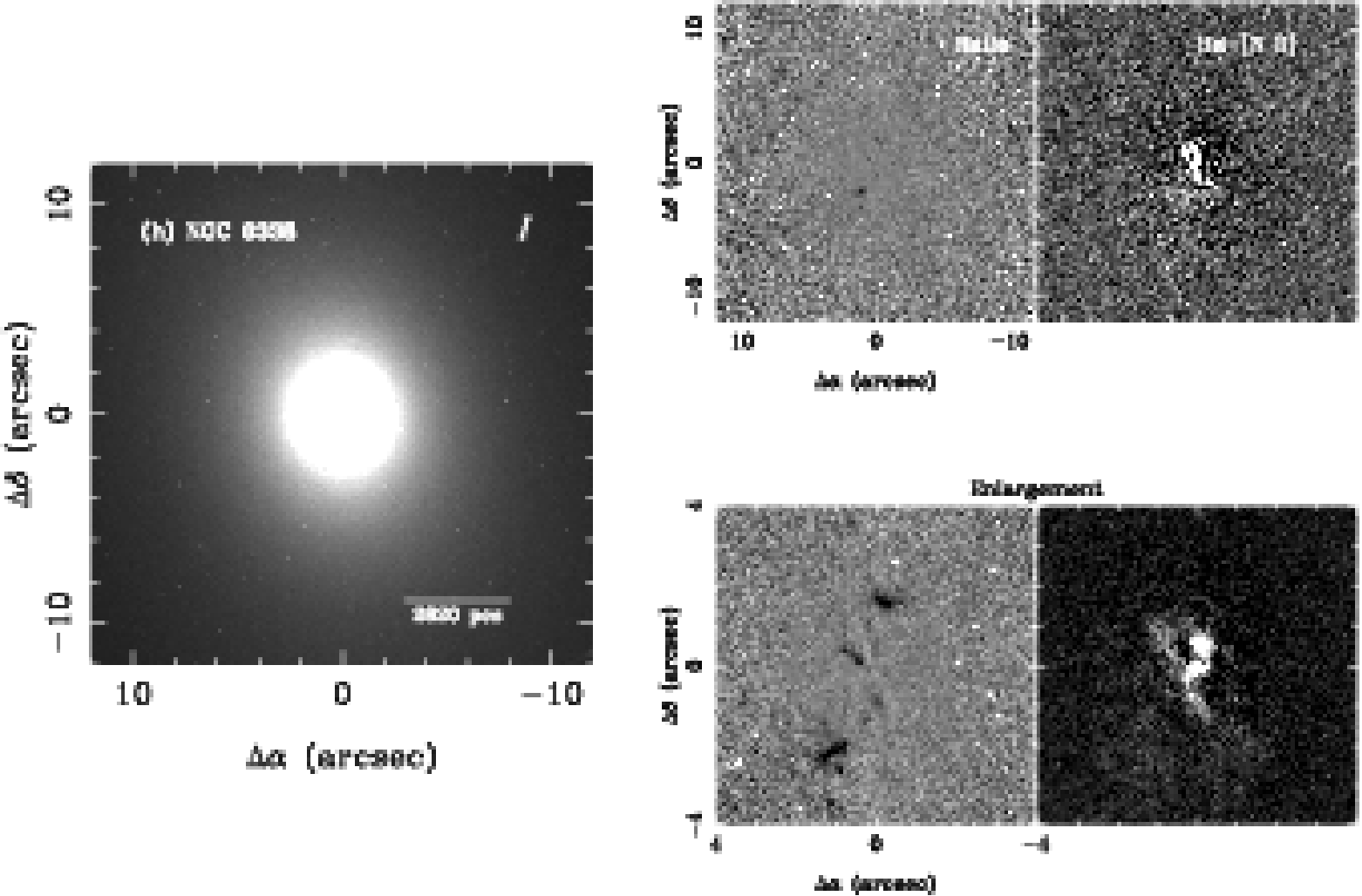}
\includegraphics[height=\textwidth,angle=270]{./martel.fig1i.ps}
\caption{{\em Continued}}
\end{figure*}

{\ \ \ } \\
\noindent
{\em NGC 7720 --- Fig.~1(i)~:} This system in Abell~2634 consists of
two galaxies separated by $12\arcsec$.  We are interested in the
southern galaxy of the pair~: it possesses a nearly face-on nuclear
dust disk and the radio source 3C~465 appears associated with it
(Venturi et al. 1995).  An H$\alpha$+[N~II] image of this galaxy as
well as an HST/WFPC2 $R$-band image of the pair are presented in
Martel et al. (2000). Our deeper ACS frames show more detail in the
structure of the disk. Also, the gas distribution is strongly peaked
on the nucleus and does not appear to extend beyond the edges of the
disk.

\begin{figure*}
\centerline{
\includegraphics[scale=0.28,angle=0]{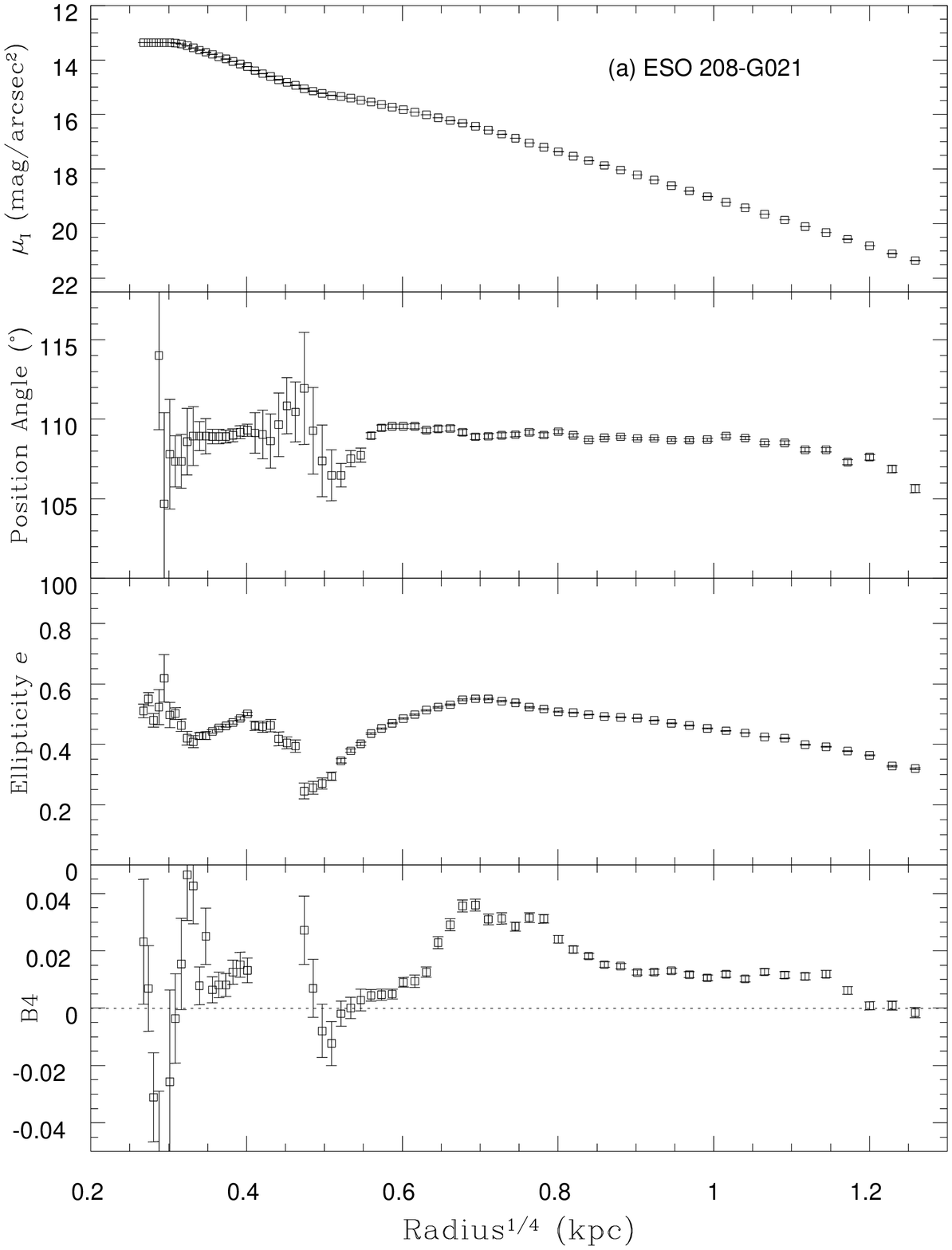}
\includegraphics[scale=0.28,angle=0]{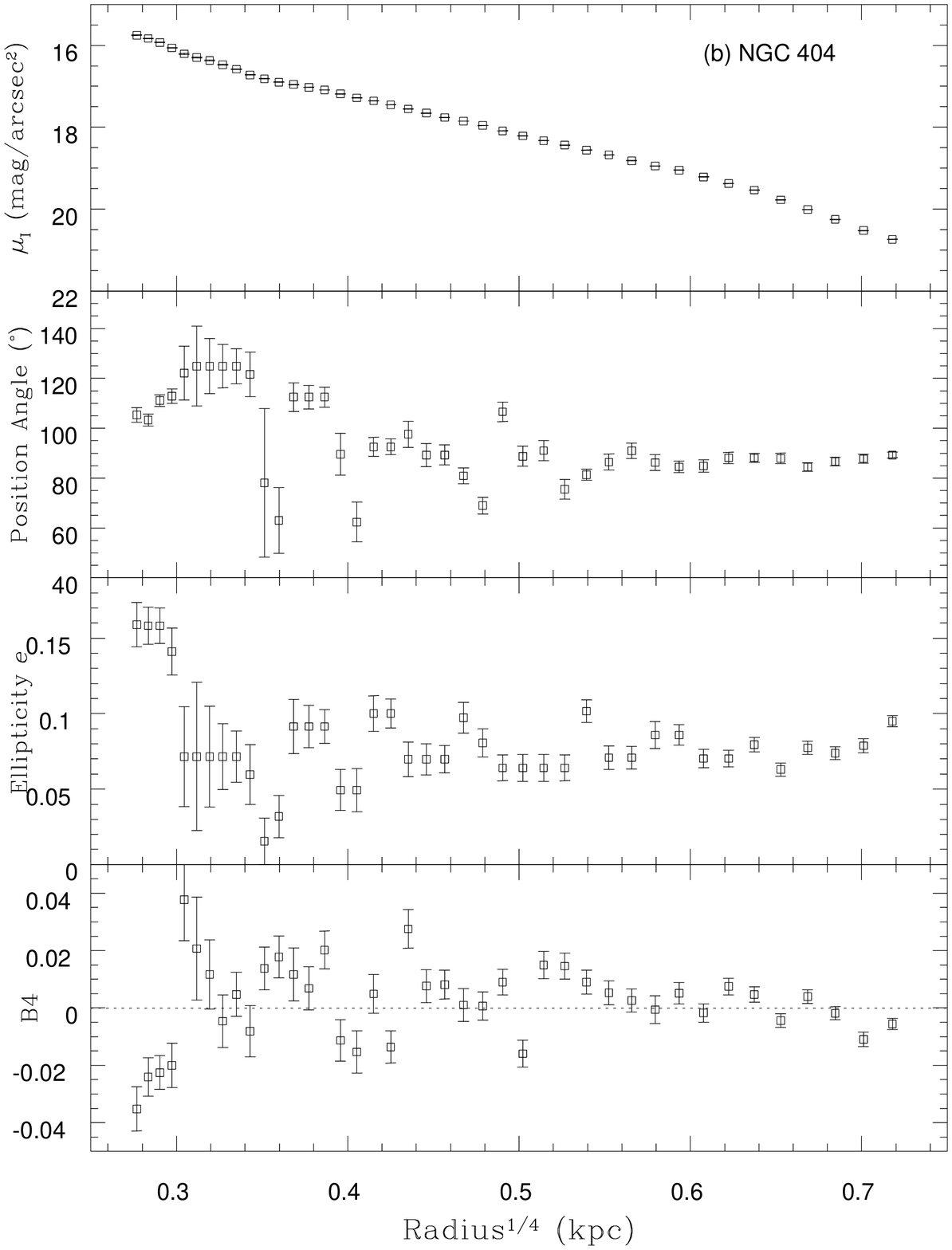}
\includegraphics[scale=0.28,angle=0]{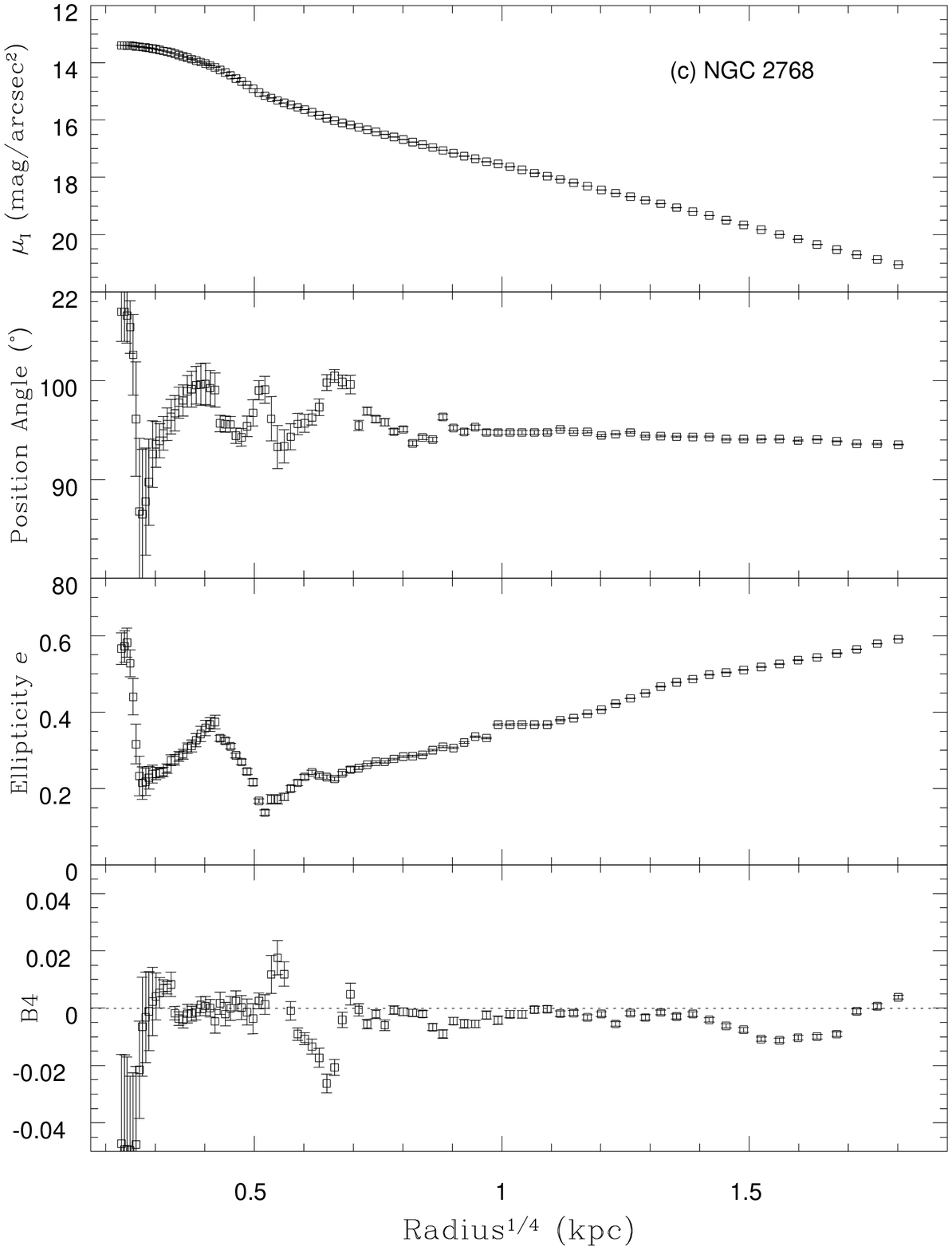}
}
\centerline{
\includegraphics[scale=0.28,angle=0]{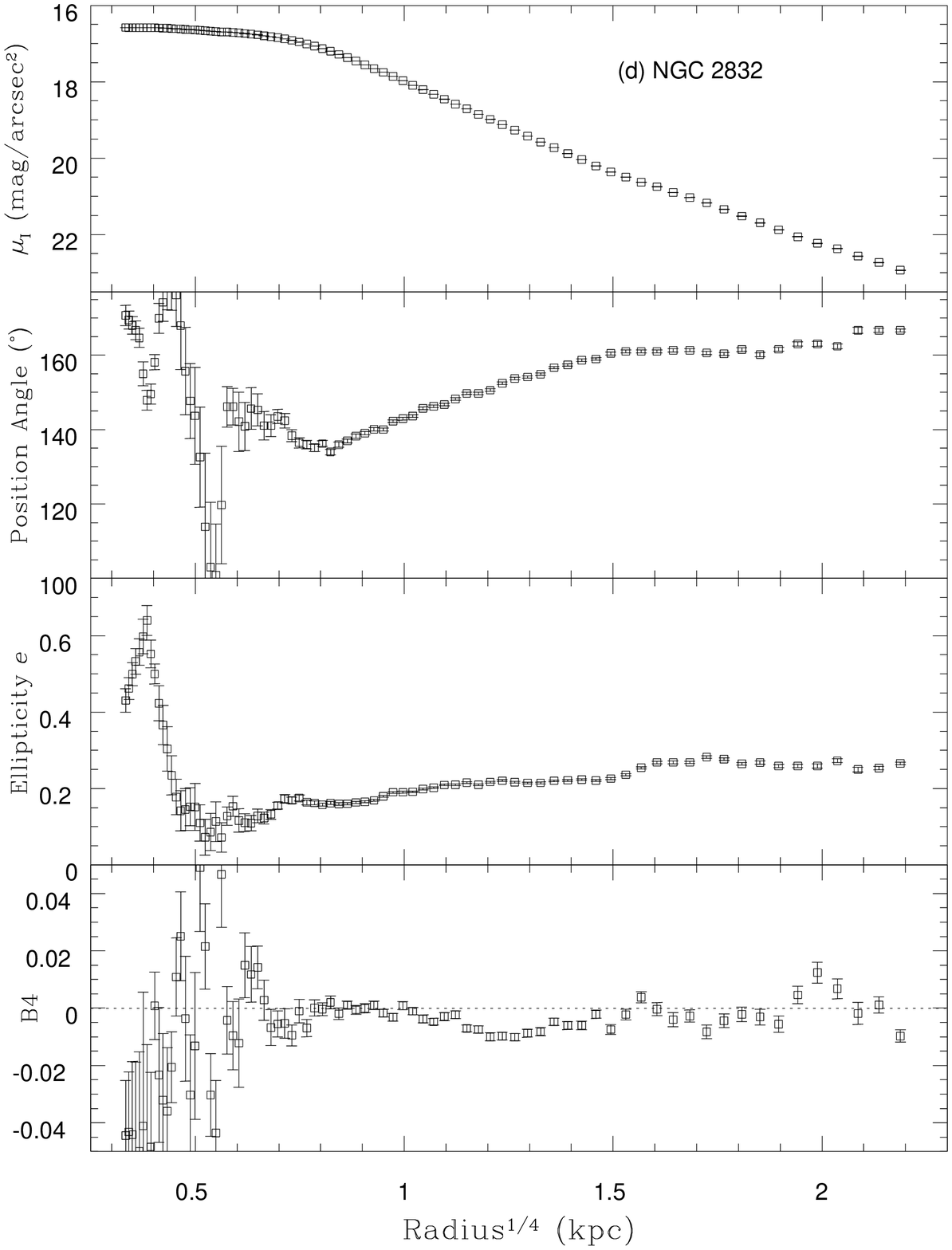}
\includegraphics[scale=0.28,angle=0]{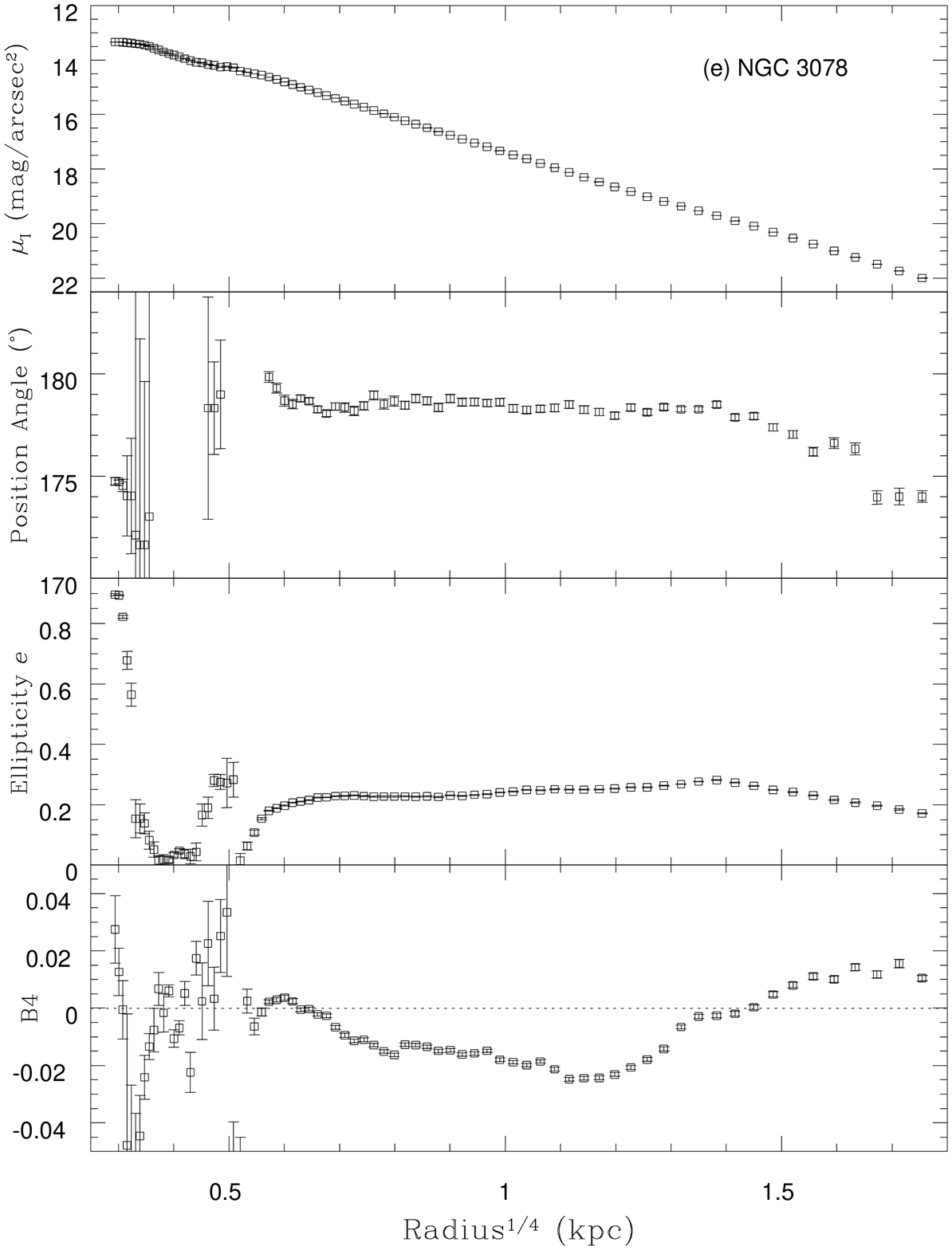}
\includegraphics[scale=0.28,angle=0]{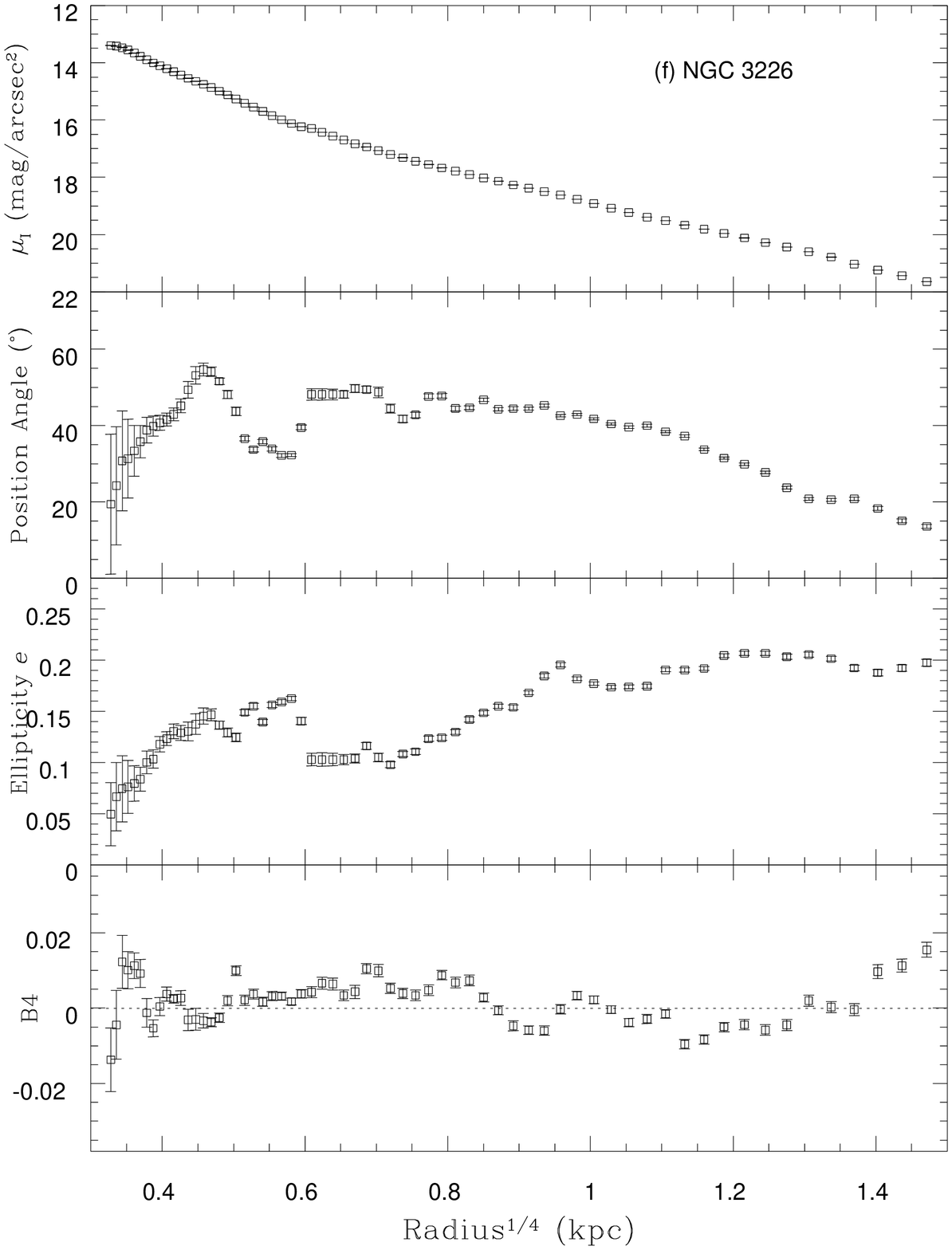}
}
\centerline{
\includegraphics[scale=0.28,angle=0]{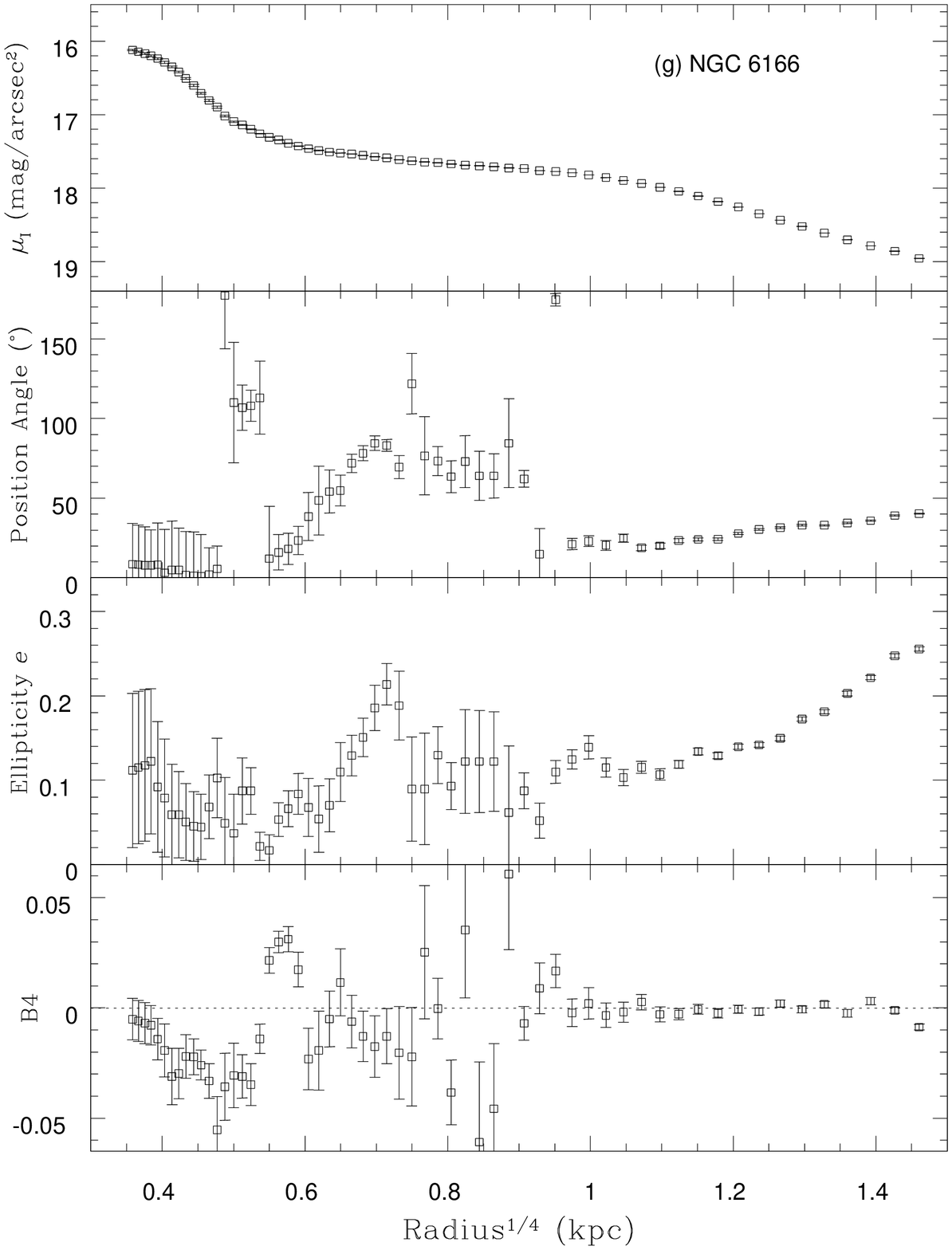}
\includegraphics[scale=0.28,angle=0]{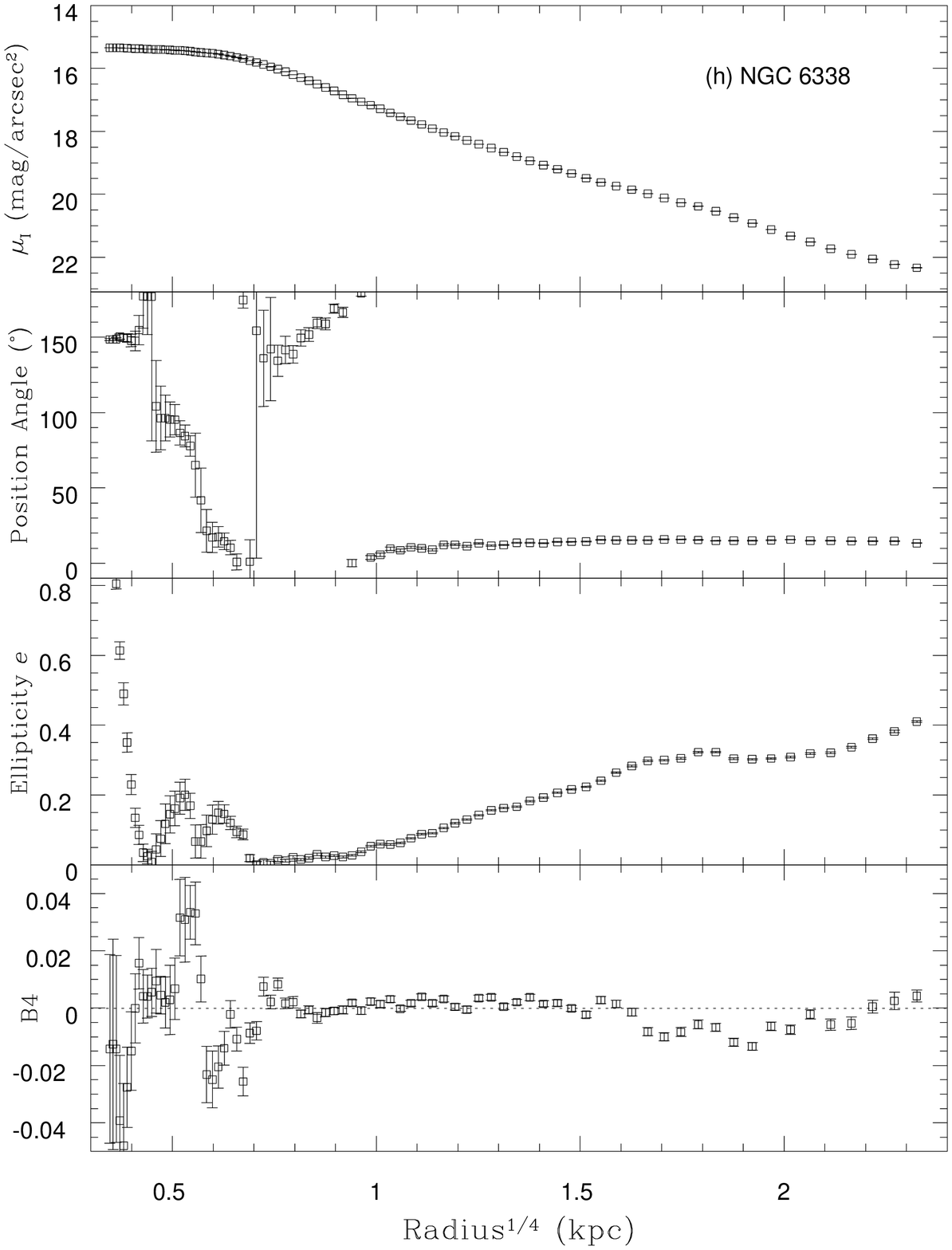}
\includegraphics[scale=0.28,angle=0]{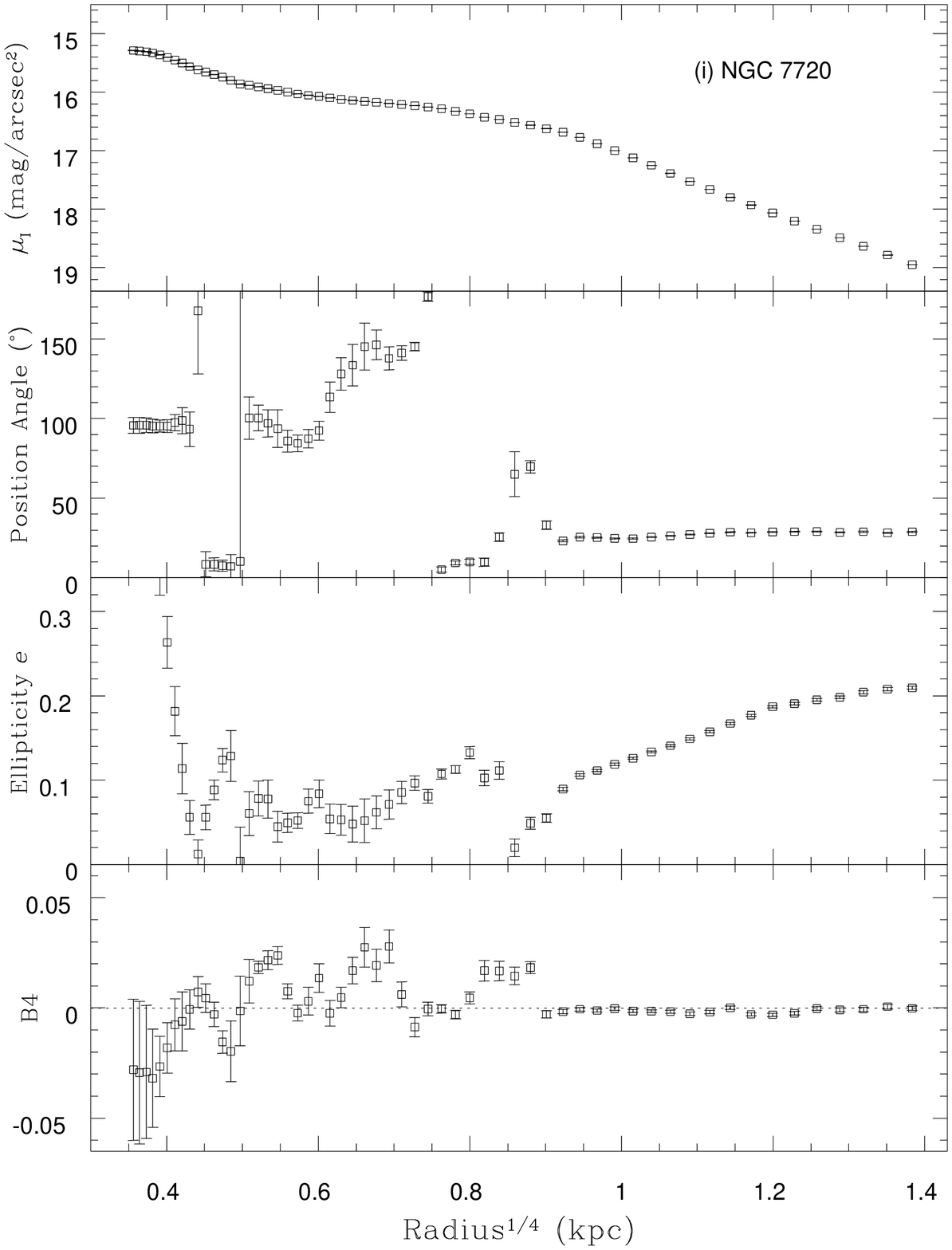}
}
\figcaption[martel.fig2.eps]{The surface brightness profile ($\mu$),
the position angle (PA), measured from north through east, the
ellipticity ($e$), and the harmonic amplitude ${\rm B}4=\cos(4\theta)$
are plotted for each galaxy of the sample.}
\end{figure*}

\subsection{Mass of the Dust and Ionized Gas}

   The measured H$\alpha$+[N~II] fluxes corrected for the redshift and
Galactic extinction are listed in cols~(2) and (3) of Table~3.  The
line flux of NGC~3078 is very uncertain and is not tabulated.  Our
deep, high-resolution ionization maps permit us to estimate the mass
$M_g$ of the ionized gas in each galaxy by making reasonable
assumptions about the volume that it occupies.  To do so, the pure
H$\alpha$ luminosity is needed, so our measurements must be corrected
for contamination by the [N~II]$\lambda\lambda6548,6583$ doublet.

   Since the dominant excitation mechanism may be different in each
galaxy, possibly resulting in a wide range of [N~II]/H$\alpha$, we
avoid using a single, generic line ratio for all objects.  A
literature search yielded [N~II]/H$\alpha$ for all our targets except
NGC~6338.  The ratios, as well as the relevant references, are given
in cols~(6) and (7) of Table~3.  Because of the lack of spatially
resolved spectroscopic information, these values are assumed to be
uniform throughout the galaxies. The final corrected H$\alpha$ fluxes
and luminosities are given in cols~(4) and (5).  The H$\alpha$
luminosity of the two radio-loud sources, NGC~6166 and NGC~7720, is
more than an order magnitude greater than that of the other ``normal''
early-type galaxies, as found in previous work (e.g. Baum \& Heckman
1989a).

   The mass of ionized material is~:

\begin{equation}
M_g=f\,V\,(n_p\,m_p+n_{\rm He}\,m_{\rm He})
\end{equation}

\noindent
where for a 10\% helium abundance~: 

\begin{equation}
n_e=(n_p+1.5\,n_{\rm He})\ {\rm and}\ n_{\rm He}=0.1\,n_p  
\end{equation}

\noindent
and from Osterbrock (1989)~:

\begin{equation}
n_e^2=\frac{1.15\,L_{H\alpha}}{f\,V\,\alpha_{H\alpha}^{eff}\,h\,\nu_{H\alpha}}
\end{equation}
where $L_{H\alpha}$ is the line luminosity in ergs~sec$^{-1}$, $V$ is
the volume occupied by the emitting gas, $f$ is the volume filling
factor, $\alpha_{H\alpha}^{eff}=1.17\times10^{-13}$~cm$^3$
sec$^{-1}$ at $T=10^4$~K for Case~B recombination, $h$ is the Planck
constant, and $\nu_{H\alpha}$ is the frequency of H$\alpha$.

   The computation of the volumes occupied by the line emitting gas is
highly uncertain due to the lack of three dimensional information.  We
assume that the volumes of the filaments are cylindrical and that the
ionized gas possesses a disk geometry in ESO~208-G021 and NGC~7720, is
conical in NGC~3226, and is spherical in the clumps of NGC~2832 and
NGC~6166.  The filling factor can be derived for the three galaxies
NGC~404, NGC~2768, and NGC~3226, given the electron densities
calculated by Ho, Filippenko, \& Sargent (1997) using the [S~II]
doublet~: $f=0.052, 0.007, 0.006$, respectively.  For the remaining
galaxies, we simply use the average $<f>\approx0.02$.  The ionized gas
masses are listed in Table~4. Although they span a wide range,
$M_g\approx7\times10^2-3\times10^6$~$M_\odot$, these are typical for
early-type galaxies (e.g. Goudfrooij et al. 1994).

   For completeness, we also calculate the mass $M_d$ of the dust in
the galaxies. This quantity is usually estimated from the optical
extinction (e.g. van Dokkum \& Franx 1995) or from the far-infrared
(FIR) emission.  Goudfrooij \& de Jong (1995) point out that the dust
masses derived from the optical extinction are significantly lower (by
a factor of $\approx8$) from those computed from the FIR flux
densities. They attribute this discrepancy to the presence of a
diffuse and uniform dusty component spread throughout the galaxy which
goes undetected in the optical imaging.  The differential extinction
from this dust, perhaps combined with metallicity effects, might
contribute to the observed color gradients in early-type galaxies
(e.g.  Silva \& Wise 1996).  But such diffuse dust may not be
necessary to account for the missing mass if a model more
sophisticated than a screen is used to calculate the extinctions.  For
example, in modeling the nuclear dusty disk of NGC~4261 with a
sandwich model that included scattering into the line-of-sight, Martel
et al. (2000) found that the mass of the disk is about an order of
magnitude greater than the mass calculated with a foreground screen
model.

   Here, to insure that we include all forms of the dust, from diffuse
to clumpy, we use the flux densities measured by the {\em Infrared
Astronomy Satellite} (IRAS).  This method provides a good estimate of
the dust mass for temperatures over which IRAS is sensitive,
$T_d\gtrsim25$~K below $100~\mu$m, but does not take into account
colder dust that radiates at longer wavelengths.  Following Goudfrooij
\& de Jong (1995) and Tran et al. (2001), the mass is given by~:

\begin{displaymath}
M_d =
5.1\times10^{-11}\,S_\nu\,D^2\,\lambda_{\mu}^4\,e^{1.44\times10^4/(\lambda_\mu\,T_d)}\:M_\odot
\end{displaymath}
where $S_\nu$ is the IRAS flux density in mJy, $D$ is the distance in
Mpc, $\lambda_\mu$ is the wavelength in microns, and $T_d$ is the dust
temperature in Kelvins.  The IRAS flux densities are taken from Knapp
et al. (1989) as tabulated in the NASA Extragalactic Database (NED).
The temperature of the dust, $T_d$, is evaluated from the
$S_{60}/S_{100}$ ratios in Kwan \& Xie (1992). The resultant dust
masses are given in Table~4. For the three ``normal'' galaxies
(ESO~208-G021, NGC~404, NGC~2768), they fall in the range
$10^3-10^5$~$M_\odot$, in accord with the results of larger surveys
(e.g. Tran et al. 2001). The dust masses of the two radio-loud
sources, NGC~6166 and NGC~7720, are larger, $10^6-10^7$~$M_\odot$.

\begin{deluxetable}{l c c}
\tablewidth{0pt}
\tablenum{4}
\tablecolumns{3}
\tablecaption{Ionized Gas and Dust Masses}
\tablehead{
\colhead{Galaxy} & 
\colhead{Gas Mass ($M_\odot$)} &
\colhead{Dust Mass ($M_\odot$)}}
\startdata  
ESO 208-G021 & $7.7\times10^2$   &   $2.8\times10^5$  \\
NGC 404      & $1.9\times10^3$   &   $7.4\times10^4$  \\
NGC 2768     & $2.4\times10^4$   &   $4.1\times10^5$  \\
NGC 2832     & $2.0\times10^5$   &   \nodata          \\
NGC 3078     & \nodata           &   \nodata          \\
NGC 3226     & $1.1\times10^4$   &   \nodata          \\
NGC 6166     & $3.5\times10^6$   &   $3.0\times10^7$  \\
NGC 6338     & \nodata           &   \nodata          \\
NGC 7720     & $3.8\times10^5$   &   $4.6\times10^6$  \\
\enddata
\end{deluxetable}



\section{Discussion}

   Our sparse sample was not explicitely selected for an in-depth
study of the ionization mechanisms and of the origin of gas and dust
in early-type galaxies, but even so, given the high quality and
spatial resolution of our images, we can make a few general remarks on
these matters.

\subsection{The Ionization Mechanisms}

   The emission-line nebulae in the early-type galaxies of our sample
are centrally concentrated inside radii of a few hundreds parsecs. In
two cases, NGC~6166 and NGC~6338, filaments of ionized gas are
detected much further out, at distances of $\approx8$~kpc.  But in all
cases, the bulk of the line emission comes from the central galactic
regions - the extended nebulae, when present, are significantly more
diffuse.  Large ground-based surveys (e.g. Kim 1989; Shields 1991;
Goudfrooij et al. 1994; Macchetto et al. 1996) have also found similar
gas distributions over similar spatial extents.  There are likely more
than one excitation mechanism at work in individual galaxies,
especially on different scales. In the following, we consider some of
the most popular mechanisms that have been discussed in the
literature.

   Even though the star formation rates in our galaxies are
insignificant, $(0.6-6)\times10^{-3}$~$M_\odot$~year$^{-1}$ using the
relation of Kennicutt (1998) and assuming that the observed H$\alpha$
is entirely produced by the star-forming process, photoionization by
massive, hot, young stars may still be responsible for some of the
ionized gas.  For example, although the dominant power source in
LINERs is still unclear, either a starburst or a diminutive active
nucleus, at least one of our three LINERs, NGC~404, is known to
possess a weak, circumnuclear star-forming region, based on its H~II
type spectrum (Ho, Filippenko, \& Sargent 1995) and UV and X-ray
morphologies and energy output (Maoz et al. 1995, 1996 ; Eracleous et
al. 2002).  Moreover, the optical nebulae in the galaxies appear
associated with cold dust, as expected for star-forming regions, and
in some cases, the gas and dust filaments are exactly
coincident. Compact clumps, a common signature of star formation, are
also observed, although smooth, kpc-scale filaments, such as in
NGC~6338, may point to another source of ionization.  High-resolution
imaging in the ultraviolet may help pinpoint active sites of star
formation in this sample, as for NGC~404.

   Photoionization by an active nucleus is likely dominant in NGC~6166
and NGC~7720.  These harbor powerful 3C radio sources and the bulk of
their line emission is concentrated on the nucleus, as is commonly
observed in radio-loud galaxies (Morganti, Ulrich, \& Tadhunter 1992).
In Fig.~3, we overlay the VLA map of Ge \& Owen (1994) atop our ACS
broad- and narrow-band images.  In the central $\approx15\arcsec$, the
radio morphology of 3C~338 consists of a jet structure terminated by
two lobes on each side of the core on the East-West axis.  Some of the
off-nuclear ionized structures are spatially correlated with this
radio morphology - this association may represent evidence of
jet-induced shock ionization.  In particular, the eastern
emission-line clump described in \S~4.1 is located exactly between the
core and the eastern lobe.  Also, one prominent ionized filament lies
along the southern edge of the western radio lobe and others are
detected near its tip.  In an HST/STIS ultraviolet map, Allen et
al. (2002) find no compact clumps at the location of the filaments,
suggesting that unextincted, active star formation is not very
important in these structures.  Confirmation of the excitation
mechanism in the extended structures of NGC~6166 will require deep,
spatially resolved spectroscopy or imaging in other diagnostic
emission lines.

\setcounter{figure}{2}
\begin{figure*}
\includegraphics[height=\textwidth,angle=270]{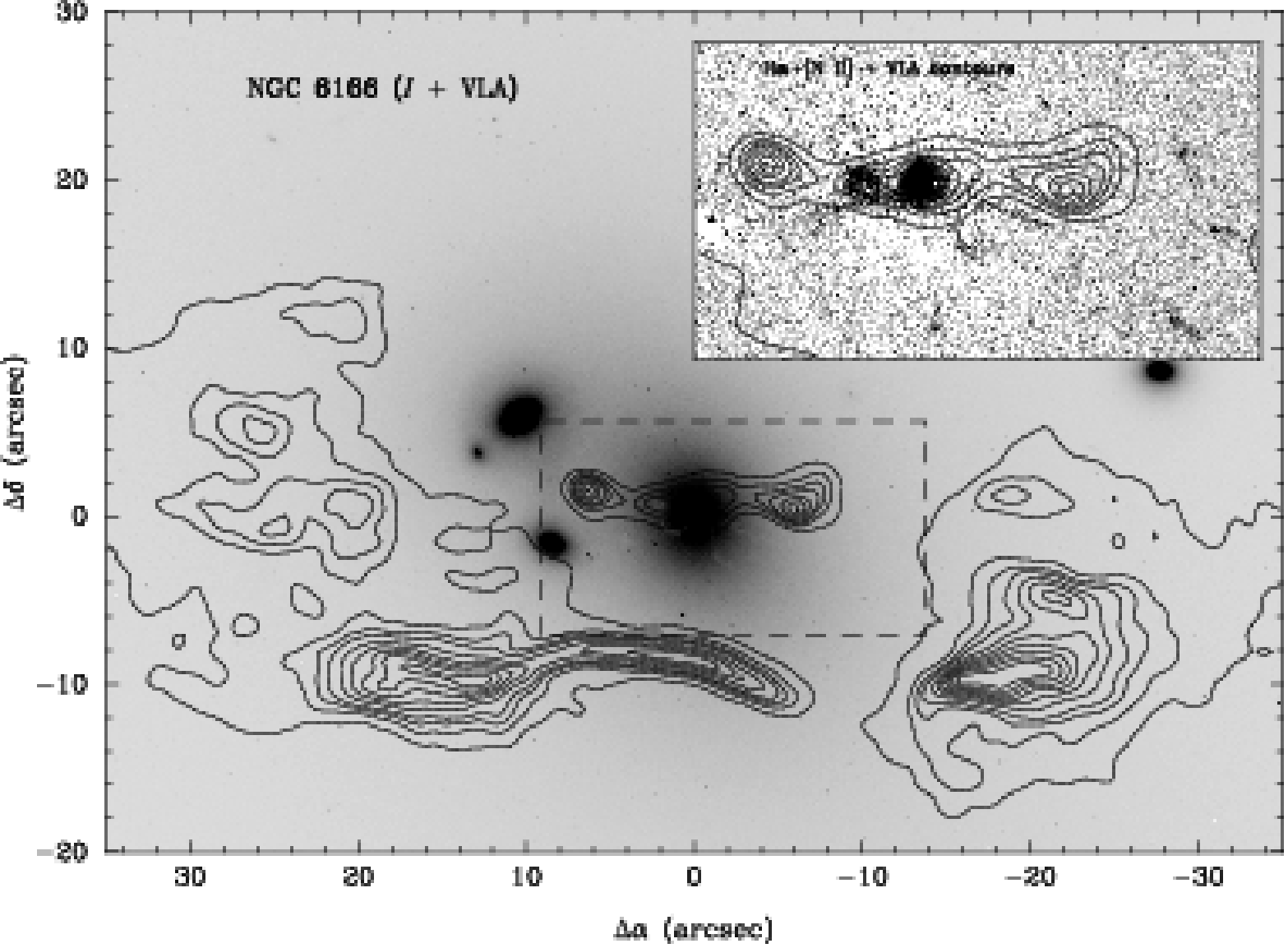}
\figcaption[martel.fig3.ps]{A VLA contour map of 3C~338 is
superimposed over the $I$-band image of NGC~6166 in the main panel and
over the H$\alpha$+[N~II] map in the inset.}
\end{figure*}

   Another possible source of ionization is thermal conduction by
electrons (e.g. Sparks, Macchetto, \& Golombek 1989).  In this
scenario, the cold gas and dust are the vestige of a merger event and
are heated by the hot, diffuse X-ray coronal halo presumed to permeate
the galaxy.  Macchetto et al. (1996) show that this mechanism predicts
H$\alpha$+[N~II] luminosities consistent with those observed in a
large number of galaxies - the line luminosities of our sample would
mostly lie at the faint end of their distribution.  Of course, the
source of X-ray emission, either hot coronal gas or discrete stellar
sources, such as low-mass X-ray binaries and supernovae, needs to be
determined for the galaxies under consideration.  From their large
catalog of pseudo-bolometric X-ray luminosities, O'Sullivan, Forbes,
\& Ponman (2001) suggest that the X-ray emission of early-type
galaxies with $L_B\lesssim10^{10}$~$L_{B,\odot}$ is dominated by
discrete sources, while the hot coronal halo dominates in the more
luminous galaxies.  Not suprisingly, in our sample, only NGC~404
satisfies the low-luminosity criterion, so thermal conduction may
prove a viable mechanism in the remaining galaxies.

   Post-Asymptotic Giant Branch (AGB) stars have also been invoked as
a source of line excitation in early-type galaxies (Trinchieri \& di
Serego Alighieri 1991; Binette et al. 1994).  These stars can produce
enough Lyman photons to account for the observed H$\alpha$
luminosities. Indeed, in their large sample of galaxies, Macchetto et
al. (1996) find that the line luminosities correlate well with the
blue luminosity inside the radius of the emitting regions, suggesting
that the excitation mechanism is tied to the old stellar population.
The model H$\alpha$ luminosities are in good agreement with their
observed values, making post-AGB ionization a plausible scenario.
Unfortunately, in our small sample, only NGC~6338 possesses
kiloparsec-scale ionized material that may be consistent with post-AGB
excitation; in the other galaxies, the line-emitting gas is strongly
concentrated in the galaxy core over scales of only a few hundred
parsecs or is dominated by some other ionization mechanism, such as a
starburst in NGC~404 and photoionization by an active nucleus in
NGC~6166 and NGC~7720. Therefore, we can not explore this excitation
mechanism in more detail.

\subsection{Origin of the Gas and Dust}

   It is generally believed that the gas and dust in early-type
galaxies is external in origin, resulting from the tidal capture a
gas-rich dwarf companion.  Evidence for this scenario comes from
spectroscopic surveys which find that the gas and stellar kinematics
are often decoupled, even counter rotating (e.g.  Bertola, Buson, \&
Zeilinger 1988, 1992; Kim 1989; Caon, Macchetto, \& Pastoriza
2000). Also, the dust and gas are usually found to be co-spatial,
suggesting that both components have the same origin.  Over timescales
of $\sim10^9$~years, the gas and dust will settle on some permitted
plane of the galaxy. Thus, any sample of early-type galaxies will
likely offer a snapshot of the dust and gas in different dynamical
states, from compact, organized, nuclear disks to extended, unsettled
filaments.  The orientation of the dust structures is usually found to
correlate with the main axes of the stellar distribution, as expected
in such a scenario (for example, see van Dokkum \& Franx (1995), de
Koff et al. (2000), and Tran et al. (2001) for radio-quiet and loud
samples).  From Fig.~4, we find that even in our limited sample, the
orientation of the H$\alpha$+[N~II] gas correlates well with either
the major or minor axis.  Since the dust and gas are essentially
co-aligned, these results also hold true for the dusty features.  One
of the exceptions is NGC~6166, whose extended line emission is best
aligned with the radio axis, as observed in other powerful radio
sources (Baum \& Heckman 1989b). Our results are therefore consistent
with past work.

\setcounter{figure}{3}
\begin{figure}
\includegraphics[height=3.2in,angle=270]{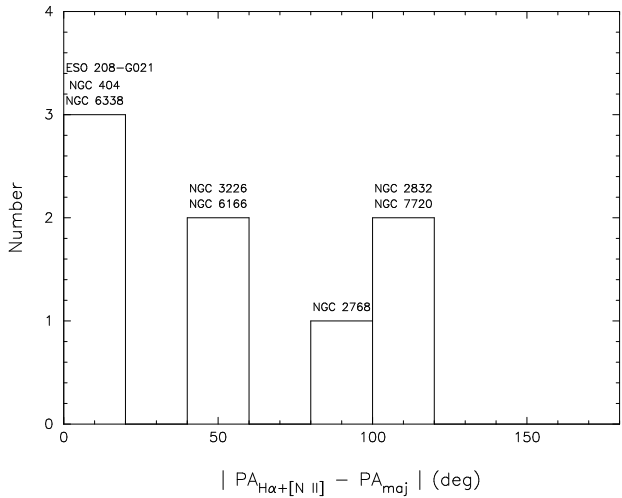}
\figcaption[martel.fig4.ps]{Angular offsets between the major axes of
the ionized gas regions and the stellar isophotes, as projected on the
sky. The ionized material is predominantly aligned along the major or
minor axis, $\Delta {\rm PA}\approx0\arcdeg$ or $\approx90\arcdeg$,
respectively.}
\end{figure}

\section{Summary}

   Resolved dust and ionized H$\alpha$+[N~II] gas are found in all
nine galaxies of our sample. In some, the dust and gas are well
organized in small compact nuclear disks while in others, they are
chaotic, filamentary, and spread throughout the galaxy. The line
emission is entirely contained within the nuclear disks of
ESO~208-G021, NGC~3078, and NGC~7720.  The three LINERs NGC~404,
NGC~2768, and NGC~3226 possess compact line-emitting regions (radii
$\lesssim100$~pc) while in the galaxies NGC~6166 and NGC~6338, the
nebulae extend to kpc scales. Different ionization mechanisms may play
a role in each galaxy, from nuclear starbursts in NGC~404 to
jet-induced shock ionization in NGC~6166.  The orientation of the
line-emitting gas is generally aligned with the minor or major axis of
the galaxies.  A detailed comparison of our new images with UV and
X-ray images will help to better understand the interaction between
the different components of the ISM in early-type galaxies.

{\ \ \ } \\
\begin{acknowledgments}
ACS was developed under NASA contract NAS 5-32865, and this research
has been supported by NASA grant NAG5-7697 and by an equipment grant
from Sun Microsystems, Inc.  The {\em Space Telescope Science
Institute} is operated by AURA Inc., under NASA contract
NAS5-26555. We are grateful to K.~Anderson, J.~McCann, S.~Busching,
A.~Framarini, S.~Barkhouser, and T.~Allen for their invaluable
contributions to the ACS project at JHU. We made use of the NASA/IPAC
Extragalactic Database (NED) which is operated by the Jet Propulsion
Laboratory, California Institute of Technology, under contract with
the National Aeronautics and Space Administration.  The FITS file of
3C~338 was downloaded from ``An Atlas of DRAGNs'' at Jodrell Bank~:
http://www.jb.man.ac.uk/atlas/.
\end{acknowledgments}

\end{document}